\documentclass[10pt, twocolumn]{article}
\usepackage[utf8]{inputenc}
\usepackage[T1]{fontenc}
\usepackage{geometry}
\usepackage{times}
\usepackage{xurl}
\usepackage{microtype}
\usepackage{amsmath, amssymb, amsfonts}
\usepackage{graphicx}
\usepackage{float}
\usepackage{caption}
\usepackage{subcaption}
\usepackage{seqsplit}
\usepackage{longtable}
\usepackage{booktabs}
\usepackage{multirow}
\usepackage{array}
\usepackage[colorlinks=true, citecolor=blue, linkcolor=blue, urlcolor=blue]{hyperref}
\usepackage[numbers]{natbib}
\usepackage{enumitem}
\setlist{nosep, leftmargin=*}
\usepackage{xcolor}
\usepackage{tabularx}
\usepackage{titlesec}
\titlespacing*{\section}{0pt}{1.5ex plus 0.5ex minus .2ex}{1.0ex plus .2ex}
\titlespacing*{\subsection}{0pt}{1.2ex plus 0.4ex minus .2ex}{0.8ex plus .2ex}
\titlespacing*{\subsubsection}{0pt}{1.0ex plus 0.3ex minus .2ex}{0.6ex plus .1ex}
\graphicspath{{figures/}}

\title{A Constrained Natural-Language Interface for Variational Multi-Physics Finite Element Simulations in FEniCS}
\author{
  \textbf{Nilay Upadhyay}\textsuperscript{1}\textbf{,}
  \textbf{Wesley F. Reinhart}\textsuperscript{2}
  \\[6pt]
  \textsuperscript{1}Department of Engineering Science and Mechanics, \\
  The Pennsylvania State University, University Park, PA 16802, USA \\[4pt]
  \textsuperscript{2}Department of Materials Science and Engineering, \\
  The Pennsylvania State University, University Park, PA 16802, USA \\[4pt]
  \texttt{nzu5027@psu.edu, reinhart@psu.edu}
}

\date{}

\begin{document}
\twocolumn[
  \begin{@twocolumnfalse}
    \maketitle
    \begin{abstract}
    \normalsize
    \noindent
Large language models can reduce the manual effort required to set up finite element simulations, but they also introduce a reliability problem when generated solver code lies on the critical path. We present a constrained natural-language interface for multi-physics finite element analysis in which the LLM is deliberately limited to front-end tasks. It parses a user prompt into a structured JSON specification, generates Gmsh code only for non-catalog geometries, and uses retry feedback for those two stages. It never writes FEniCS solver templates, never derives weak forms, and never writes the numerical solver core. A deterministic dispatcher maps the validated specification to one of five human-written FEniCS/UFL templates: linear elasticity, hyperelasticity, elastoplasticity, thermo-mechanical coupling, and phase-field fracture. We validate this deterministic template layer against analytical solutions and published benchmarks in two and three dimensions. Smooth cases reach sub-percent agreement on adequate meshes, while the harder nonlinear cases reach the 2--5\% range. We also evaluate the LLM-facing front end directly. In a 15-prompt parser benchmark, first-pass valid parses were obtained for 9 cases (60.0\%), and all remaining cases were repaired after retry, giving a final valid parse rate of 100.0\%, 100.0\% problem-class accuracy, and 97.1\% field-extraction accuracy. In a 10-case custom geometry benchmark routed through the real LLM-to-Gmsh path, first-pass success was 90.0\% and the final success rate was 90.0\% as well, with one unrecovered invalid-geometry failure. These results show that the parser is effective on this benchmark under a repair-based design and that the current constrained prompt and validation design is effective on this custom-geometry benchmark. As an end-to-end demonstration, the system generates and analyzes a three-dimensional elastoplastic L-bracket with a fillet and bolt hole from one natural-language prompt. The contribution is a measured architecture for natural-language-driven variational simulation, not open-ended autonomous code generation.
\end{abstract}

\vspace{1em}
\end{@twocolumnfalse}
]

\section{Introduction}\label{sec:introduction}

Finite element analysis underpins many quantitative engineering decisions in structural, mechanical, and materials design \cite{hughes2012finite, zienkiewicz_taylor_fem, bathe_fem_procedures, belytschko_nonlinear_2014, wriggers_nonlinear_2008}. Despite decades of progress in solver technology and software, the workflow around the solver remains labor intensive. An engineer translates a physical problem into a geometry, selects a constitutive law, prescribes boundary conditions, generates a mesh \cite{geuzaine_remacle_gmsh}, configures a solver \cite{petsc_users_manual}, and inspects the output. Each step requires judgment and can produce a wrong result without an obvious error message. Even with high-level libraries such as FEniCS \cite{logg2012automated, alnaes_ufl_2014}, which remove much of the low-level assembly work, the analyst still has to map the engineering description into a correct variational problem.

Recent efforts to accelerate computational mechanics and PDE simulation have heavily explored data-driven surrogate models, including Physics-Informed Neural Networks (PINNs) \cite{raissi2019physics}, neural operators \cite{lu2021learning, li2021fourier}, and graph neural networks for mesh-based simulations \cite{pfaff2021learning, sanchez2020learning}. While these approaches attempt to replace or accelerate the numerical solver itself, large language models offer a different approach: acting as an interface layer surrounding a traditional solver. Recent work has used LLMs to drive scientific computing tasks through theoretical specification frameworks coupled to deterministic solvers \cite{simulability2026}, multi-agent systems that synthesize numerical solvers \cite{autonumerics2026}, and pipelines that convert physical component descriptions or images into engineering reports \cite{perception2026}. Related efforts have appeared in chemistry \cite{chemcrow2023} and computational fluid dynamics with OpenFOAM \cite{openfoamgpt, metaopenfoam, foamagent}. These systems differ in scope, but they raise the same design question: where should the LLM sit in the computational stack?

This question determines the failure modes of the system. If the LLM generates solver-specific input files, the system inherits the limits of that input schema. If the LLM generates FEniCS code directly, then LLM-generated solver code becomes part of the numerical method. Hallucinated weak forms, wrong boundary conditions, or invalid solver logic can then pass into the simulation unless caught later. The abstraction boundary is therefore a mechanics and reliability decision, not only a software design decision.

The phase-field model for brittle fracture \cite{miehe2010thermodynamically, bourdin_francfort_marigo, ambrosio_tortorelli, borden2012phase, kuhn2010continuum} illustrates the issue. In an input-card system, supporting phase-field fracture requires the solver to expose a damage degree of freedom, a history field with irreversibility, a staggered or monolithic coupling between displacement and damage, and output cards for the damage variable. If the schema does not expose these objects, the user cannot run the model. In a UFL-based system, phase-field fracture is expressed as a degraded elastic energy plus a damage evolution equation derived from the Ambrosio--Tortorelli regularization. The assembler and solver see weak forms. The same structure also supports sequential thermo-mechanical coupling and history-dependent models such as J2 plasticity \cite{simo_hughes_plasticity, de2008computational}. This is why variational-level automation is attractive for multi-physics mechanics.

Some initial LLM-driven FEA systems followed the input-card route. FeaGPT \cite{feagpt} drives CalculiX through generated input files and demonstrates a 432-configuration parametric study of NACA airfoils, but the published results are restricted to linear elasticity. AutoFEA \cite{autofea} grounds LLM-generated CalculiX cards in a graph neural network retrieval system over a curated corpus, which improves reliability on covered cases but inherits the same physics limits outside that corpus. Multi-agent systems for two-dimensional frame analysis through OpenSees \cite{frame_agent_2025, structural_llm_2025} and benchmark studies against COMSOL \cite{feabench} occupy related territory. The COMSOL benchmark reported that the best LLM agent strategy generated executable API calls 88\% of the time without producing a fully correct solution on any problem in the suite. These systems show the value of natural-language interfaces, but they also show that solver-specific syntax can cap the physics and that executable code is not the same as a correct analysis.

A second group of systems moved to FEniCS. MCP-SIM \cite{mcpsim} wraps FEniCS code generation in a memory-coordinated multi-agent loop with clarification, error diagnosis, and revision. ALL-FEM \cite{allfem} fine-tunes open-weight language models from 3 to 120 billion parameters on more than a thousand verified FEniCS scripts and uses Formulator, Coder, Executor, and Corrector agents. Its best configuration reaches a 71.79\% code-level success rate across 39 problems spanning solid mechanics, fluids, and fluid-structure interaction. Studies of multi-agent FEA frameworks have also reported failure modes such as affirmation bias and premature consensus among reviewer agents \cite{tian_zhang_collab2025}. These FEniCS-based systems use the right mathematical abstraction, but they still place LLM-generated FEniCS code on the solver path.

We take a different position. If hallucinated FEniCS code is a central failure mode, the safest response is to remove the LLM from the FEniCS code path. In the system presented here, an off-the-shelf LLM parses a natural-language description into a structured JSON specification with a small, closed set of fields. A deterministic dispatch then routes that specification to one of five human-written UFL templates: linear elasticity, hyperelasticity with a compressible Neo-Hookean model \cite{ogden_nonlinear_elasticity, holzapfel2000nonlinear, bonet2008nonlinear}, elastoplasticity with J2 return mapping and linear isotropic hardening \cite{simo_hughes_plasticity, de2008computational}, sequential thermo-mechanical coupling, and phase-field fracture under the AT2 model \cite{miehe2010thermodynamically, ambrosio_tortorelli}. The LLM never writes FEniCS code. It never derives a weak form. It never edits the solver template. This is an intentional constraint.

The LLM is used in only three places. First, it maps the user prompt to a structured specification. Second, it generates Gmsh code when the geometry is not covered by the deterministic geometry catalog. Third, it receives error messages and retries those two front-end steps when validation fails. All numerically sensitive content is handled by deterministic, inspectable code. This design has a cost: adding a sixth physics class requires writing and reviewing a new template. It also has a benefit: the mathematical form solved by FEniCS is not produced by the LLM.

This paper evaluates both sides of that design. The deterministic template layer is validated against analytical solutions and published benchmarks across five physics classes. The LLM-facing interface layer is evaluated separately with two front-end benchmarks: a parser robustness benchmark and a custom geometry-generation benchmark. The parser benchmark measures whether natural-language prompts become valid structured specifications under a repair-and-validation loop. The geometry benchmark measures the real custom LLM-to-Gmsh path and its failure modes. This distinction matters. Without the front-end benchmark, the paper would only show that the FEniCS templates work. Without the mechanics validation, the paper would only show that the interface produces plausible inputs. The contribution is the combination: a constrained language interface over deterministic variational templates, with both layers measured.

We also demonstrate the full pipeline on a three-dimensional steel L-bracket with a fillet and bolt hole. This geometry is not in the parametric catalog and must be generated through the LLM-to-Gmsh path. The elastoplastic analysis with forty load steps completes in 87.5 seconds from one natural-language prompt without manual editing. This case is a successful end-to-end demonstration, but it is not a substitute for aggregate reliability. The separate geometry benchmark gives the reliability result for this path under the current prompt and validation design.

The remainder of this paper is organized as follows. Section~\ref{sec:architecture} describes the seven-stage pipeline architecture and defines the LLM boundary. Section~\ref{sec:mechanics_validation} reports deterministic mechanics validation across five physics classes. Section~\ref{sec:frontend_benchmarks} reports the front-end parser and geometry benchmarks. Section~\ref{sec:parametric_studies} covers parametric studies and gradient-free optimization. Section~\ref{sec:lbracket} presents the L-bracket demonstration. Section~\ref{sec:performance} reports computational performance. Section~\ref{sec:robustness} discusses reproducibility, prompt sensitivity, failure modes, and limitations. Section~\ref{sec:discussion} compares the design with recent LLM-FEA systems. Section~\ref{sec:conclusion} concludes.

\begin{figure*}[h!]
\centering
\includegraphics[width=0.95\textwidth]{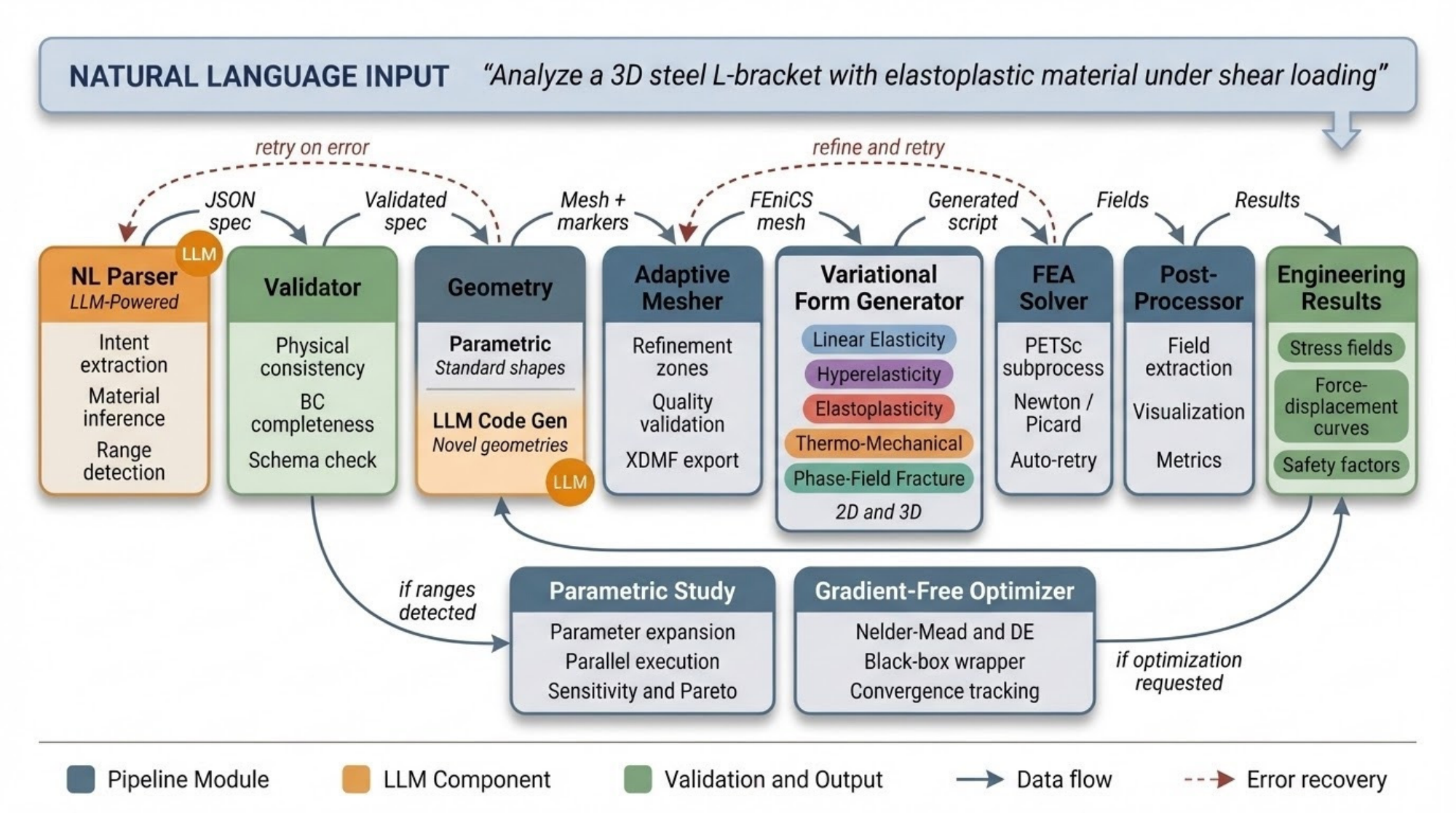}
\caption{Architecture of the FEniCS agent. A natural-language input is parsed into a structured specification, validated for physical consistency, and routed through single-analysis, parametric-study, or optimization modes. The LLM is used for parsing, custom Gmsh geometry generation, and retry feedback for those two stages. The solver path is deterministic: validated specifications are mapped to human-written FEniCS/UFL templates. Failed solves trigger deterministic retries with refined meshes or adjusted solver parameters.}
\label{fig:architecture}
\end{figure*}

\begin{table*}[t!]
\centering
\caption{Comparison of LLM-driven FEA systems. Input-card systems are limited by solver schemas. FEniCS-based systems operate at the variational form level, but differ in whether the LLM writes solver code. This work keeps the LLM out of the FEniCS solver path and evaluates the LLM-facing front end separately.}
\label{tab:comparison}
\small
\renewcommand{\arraystretch}{1.2}
\begin{tabularx}{\textwidth}{l l >{\raggedright\arraybackslash}X >{\raggedright\arraybackslash}X >{\raggedright\arraybackslash}X >{\raggedright\arraybackslash}X}
\toprule
\textbf{System} & \textbf{Solver} & \textbf{LLM role in solver code} & \textbf{Physics} & \textbf{Geometry} & \textbf{Optimization support} \\
\midrule
FeaGPT & CalculiX & Generates input cards & Linear elasticity & FreeCAD & Parametric sweep \\
AutoFEA & CalculiX & Generates input cards (GNN-retrieval grounded) & Linear elasticity & None & None \\
Frame agent & OpenSees & Generates Python code & 2D linear frames & None & None \\
FEABench & COMSOL & Generates API calls & Linear benchmark & API-driven & None \\
MCP-SIM & FEniCS & Generates FEniCS code (multi-agent correction) & Mixed 12-task benchmark & Limited & None \\
ALL-FEM & FEniCS & Generates FEniCS code (fine-tuned LLMs) & Solid, fluid, multiphysics & Limited & None \\
\textbf{This work} & \textbf{FEniCS} & \textbf{None (deterministic templates)} & \textbf{Linear, hyperelastic, plastic, thermal, fracture} & \textbf{Parametric + LLM-generated Gmsh} & \textbf{Gradient-free design optimization} \\
\bottomrule
\end{tabularx}
\end{table*}

\section{System Architecture}\label{sec:architecture}

\subsection{Pipeline Overview}\label{sec:pipeline_overview}

The agent transforms a natural-language description into a complete simulation through seven stages, shown in Figure~\ref{fig:architecture}. A parser converts the text into a structured specification, a validator checks the specification for physical consistency, a geometry module produces a meshable shape, an adaptive mesher generates a discretization, a template assembler emits a FEniCS script, a solver executes that script, and a post-processor extracts results. A mode router placed after validation sends the specification to a single analysis, parametric-study, or optimization pipeline.

The stages communicate through one JSON schema. Each module reads and writes the same dictionary structure. This makes it possible to log a run, reproduce it from a saved specification, and replace the LLM backend without changing the solver layer. It also makes the stages independently testable. A hand-written specification can be injected at any stage to isolate whether an error comes from parsing, geometry, meshing, template assembly, solution, or post-processing. 

\subsection{Natural Language to Structured Specification}\label{sec:nl_to_spec}

The parser uses an LLM to extract a structured specification from free-form engineering text. Its system prompt contains the JSON schema, inference rules for common engineering defaults, and examples for each supported problem class. The model is run with temperature zero so that repeated parses of the same input are stable. The Supplementary Information provides an example of the JSON schema from one of our runs.

The inference rules are intentionally limited. A description that mentions ``aerospace bracket'' without material data may default to aluminum 7075-T6. A description that mentions ``rubber'' or ``elastomer'' selects a hyperelastic problem class with a Neo-Hookean model. A description that mentions ``crack'' or ``damage propagation'' selects phase-field fracture. These rules fill common omissions, but they do not replace validation. Under-specified prompts can still be rejected. Any inferred defaults are written into the structured specification and checked by the validator.

After parsing, the validator checks that the specification is physically meaningful. It enforces positive material constants, valid Poisson ratios, boundary conditions that suppress rigid body motion, and compatibility between material and constitutive model. If a check fails, the validator creates a targeted error message and re-prompts the parser with the original description and the specific issue. Up to three correction attempts are allowed before the system reports an unrecoverable parsing failure.

\subsection{Geometry Generation}\label{sec:geometry_generation}

The geometry stage has two paths. Standard shapes such as rectangles, plates with holes, boxes, cylinders, and notched specimens are produced by deterministic generators that call the Gmsh OpenCASCADE API directly. These generators are used for the validation suite.

When the parser sets the shape type to custom, the geometry stage uses LLM-generated Gmsh code. The prompt contains the geometry description, the required boundary names, lightweight boundary-condition or load-role hints for those names, API references for relevant Gmsh functions, and worked examples. The model is instructed to identify candidate faces first, collect matching surface tags into lists, and create exactly one physical group per required named boundary only after confirming that matching surfaces exist. It is also instructed not to introduce extra geometric features that were not requested; fillets and chamfers are only allowed when the prompt explicitly asks for them.

Before execution, the code is parsed for syntax, scanned for forbidden operations such as file I/O and shell calls, and checked against narrow pre-execution guards for known bad patterns observed in failed runs. In the current implementation, these guards reject uses of \texttt{gmsh.model.occ.getBoundary}, \texttt{removePhysicalGroups()}, and unrequested \texttt{.fillet(} calls. The code then runs in a restricted execution namespace with a timeout. Between retries, the Gmsh model state is reinitialized to avoid stale state contamination. The resulting geometry is validated by checking that a positive volume exists, that all required boundary names appear in the mapping, and that each tagged surface has nonzero area. If geometry validation fails, the failure is fed back to the model for regeneration, with up to three attempts. If geometry validation succeeds but meshing fails, the pipeline allows one additional conservative regeneration with mesh-failure feedback.

\subsection{Adaptive Meshing}\label{sec:adaptive_meshing}

The mesher applies four density levels: coarse, medium, fine, and ultra-fine. Each level is defined relative to a characteristic length of the geometry. The specification can also request local refinement near load surfaces, hole boundaries, notch tips, and material interfaces. Refinement is implemented with Gmsh distance and threshold fields combined through a minimum field, so multiple refinement zones can be active at once.

For phase-field fracture, the mesher can apply local refinement near the notch or expected crack path when such a refinement region is requested. This local refinement rule exists in the backend, although the saved audit artifact did not preserve enough metadata to verify an explicit post hoc \(\ell\)-based \(h_{\max}\) criterion for the archived phase-field case. Without a suitable local refinement region, the phase-field benchmarks often converge to the wrong onset displacement.

Gmsh output is converted to FEniCS through meshio. The conversion produces an XDMF mesh file and separate XDMF files for boundary and subdomain markers. Marker integers match the boundary map produced by the geometry stage, so semantic names in the specification map directly to MeshFunction values used for boundary conditions and surface integrals.

\subsection{Geometry-to-Mesh Transfer Audit}\label{sec:mesh_audit}

To check that successful geometries remain usable by the solver, we audited the saved geometry-to-mesh handoff artifacts. As summarized in Table~\ref{tab:mesh_audit}, 15 cases were inspected and 14 reached the meshing stage. All 14 meshed cases generated a Gmsh mesh successfully, converted to XDMF successfully, and preserved all required boundary markers after conversion. No required marker was empty, and no downstream failure was traced to mesh generation, XDMF conversion, or boundary-marker loss. The only case without mesh data was the custom obround-slot geometry, which failed during geometry generation before meshing. For the saved phase-field case, the observed notch-boundary segment size was approximately 0.020 for a fracture length scale of 0.03. However, the saved artifacts did not encode an explicit required local $h_{\max}$ rule, so the phase-field mesh-size criterion could not be verified in this audit.

\begin{table}[h!]
\centering
\caption{Geometry-to-mesh transfer audit on saved benchmark artifacts. Marker preservation is evaluated only for cases that reached meshing.}
\label{tab:mesh_audit}
\small
\renewcommand{\arraystretch}{1.15}
\begin{tabular}{lr}
\toprule
Metric & Value \\
\midrule
Cases inspected & 15 \\
Cases reaching meshing & 14 \\
Gmsh mesh successes & 14 / 14 \\
XDMF conversion successes & 14 / 14 \\
Boundary marker preservation & 14 / 14 \\
Empty required boundary markers & 0 \\
Downstream mesh/marker failures & 0 \\
Cases failing before meshing & 1 \\
\midrule
Phase-field cases checked & 1 \\
Observed notch segment size & 0.020 \\
Fracture length scale & 0.030 \\
Explicit $h_{\max}$ rule verified & No \\
\bottomrule
\end{tabular}
\end{table}

\subsection{Variational Form Generation}\label{sec:variational_form}

This stage is the main architectural boundary. Prior FEniCS-based systems use an LLM, sometimes with fine-tuning or multiple agents, to generate FEniCS code from the parsed problem. Our system does not. The variational form generator is a deterministic dispatcher. It maps the \texttt{problem\_class} field of the validated specification, one of \texttt{linear\_elastic}, \texttt{hyperelastic}, \texttt{elastoplastic}, \texttt{thermo\_mechanical}, or \texttt{phase\_field}, to one human-written template. Each template is a parameterized Python function that emits a complete FEniCS script when given the specification, mesh paths, and boundary marker map. The LLM contributes nothing to this stage.

For linear elasticity, the template constructs a vector function space, defines strain and Cauchy stress tensors using Lam\'{e} parameters, and assembles the bilinear and linear forms. For hyperelasticity, the template uses a nonlinear displacement function, defines the deformation gradient and right Cauchy--Green tensor, and writes a Neo-Hookean strain energy density from which UFL derives the residual and tangent. For elastoplasticity, the template implements an initial stiffness Newton scheme around a J2 return mapping algorithm with linear isotropic hardening. The elastic stiffness is assembled once, strain is projected to a piecewise-constant tensor space, trial stresses are radially returned to the yield surface, and the residual drives the displacement correction. For sequential thermo-mechanical coupling, the template solves a thermal Poisson problem first and then solves mechanics with an initial thermal strain term. For phase-field fracture, the template implements a staggered AT2 scheme with degraded elastic energy and a history field for irreversibility.

The specification controls function spaces, loads, boundary conditions, solver settings, load stepping, and requested outputs through fields populated by the parser and checked by the validator. The generated script is self-contained and can be run independently. Adding a new physics class requires one new template and one dispatch-table entry. It does not require a fine-tuning dataset, and it does not require the LLM to learn new weak forms.

\subsection{Execution and Error Recovery}\label{sec:execution}

Each generated script runs as a separate Python subprocess. This isolates PETSc state between runs, prevents accumulated memory effects across cases, and lets the orchestrator enforce timeouts. The solver writes a results JSON file with scalar metrics and XDMF files for field outputs.

When a solve fails, the orchestrator classifies the error as a convergence failure, memory issue, mesh problem, or setup error. Convergence failures trigger a retry with a finer mesh or smaller load increment. Memory failures trigger a retry with a coarser mesh or a switch to an iterative solver. Up to three retries are allowed before the run is marked failed and reported with the diagnostic trace. This retry logic is deterministic and separate from the LLM parser retry loop.

\subsection{Parametric and Optimization Modes}\label{sec:parametric_mode}

The parametric mode is triggered when the specification contains parameter ranges. A parameter expansion routine builds the Cartesian product of the ranges and creates one fixed-value specification per configuration. Each configuration runs in its own working directory, and configurations execute in parallel using a process pool sized to the workstation. Results are aggregated into a CSV file. The analysis module then computes parameter-metric correlations, extracts Pareto-optimal designs for multi-objective cases, and produces sensitivity plots.

The optimization mode is triggered when the specification contains an optimization section. A wrapper takes design variables, fills them into a base specification, and runs the full geometry-mesh-solve-extract pipeline as a black-box objective. The wrapper is passed to scipy’s bounded scalar search, Nelder--Mead, COBYLA, or differential evolution routines \cite{virtanen_scipy_2020}, depending on the dimension, bounds, and constraints of the problem. This is gradient-free design optimization, not topology optimization. It is suitable for low-dimensional design spaces and for physics with path-dependent or staggered solves.

\begin{figure*}[h!]
\centering
\includegraphics[width=0.95\textwidth]{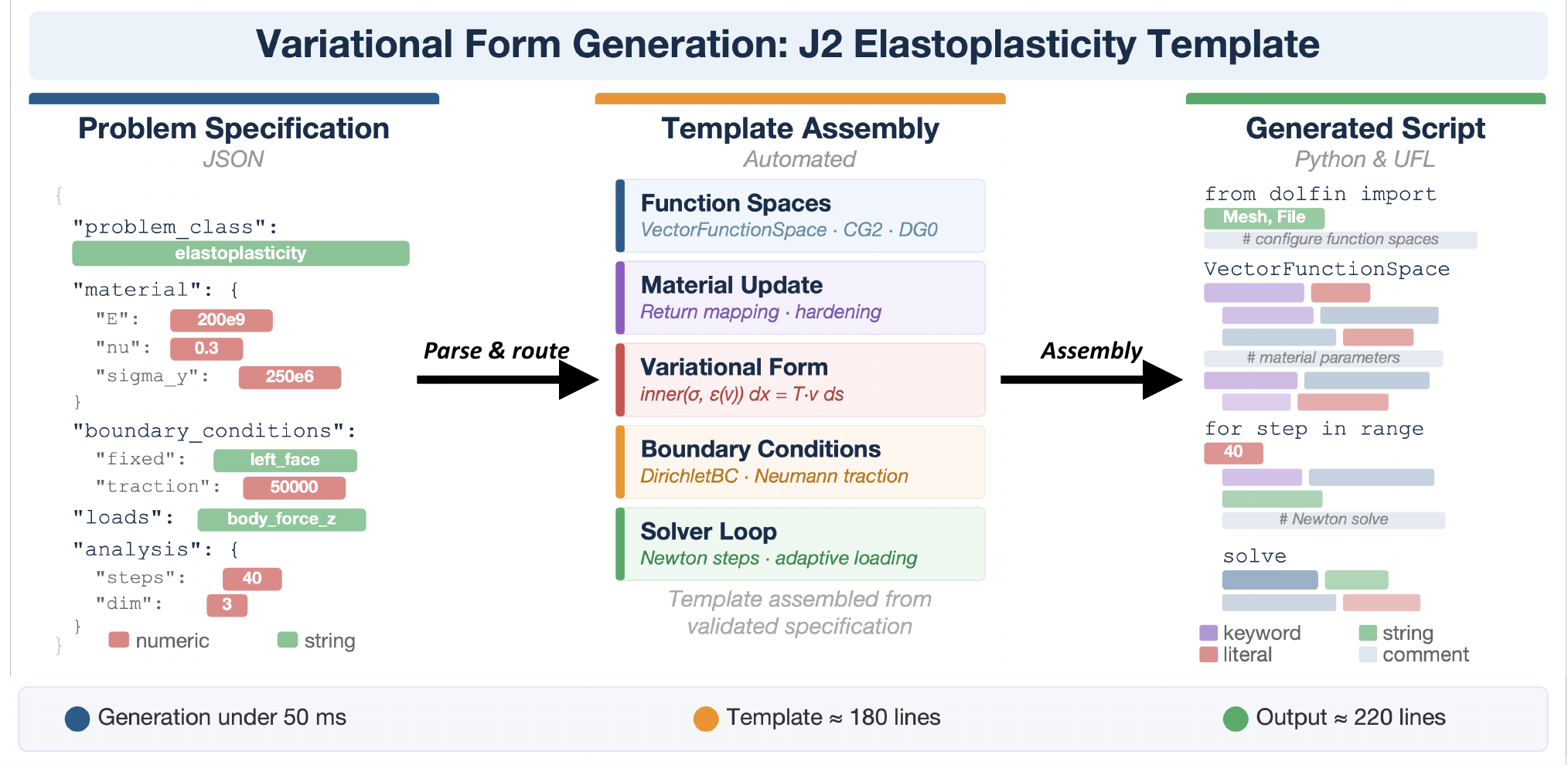}
\caption{Variational form generation for elastoplasticity. The structured specification is routed through a deterministic template that selects function spaces, the material update routine, the variational form, boundary conditions, and the solver loop. The output is a self-contained FEniCS Python script. No LLM writes the weak form or solver code.}
\label{fig:figure2}
\end{figure*}

Figure~\ref{fig:figure2} illustrates how a validated specification becomes a generated FEniCS script for elastoplasticity.

\section{Deterministic Mechanics Validation}\label{sec:mechanics_validation}

We validate the deterministic template layer on benchmarks drawn from analytical solutions, published reference results, and mesh convergence studies. Each problem class has at least one quantitative comparison against a known answer. Most cases include multiple mesh densities to separate template correctness from discretization artifacts. Where possible, the cases are standard benchmarks with independent reference values.

Errors are reported as relative percentage errors against the reference value at the comparable evaluation point: $|x_{\mathrm{FEM}} - x_{\mathrm{ref}}| / |x_{\mathrm{ref}}| \times 100\%$. For convergence studies, the reported error is the error of the scalar quantity of interest, such as tip deflection, peak stress, or fracture onset displacement. For continuous comparisons such as the Lam\'{e} radial stress profile, the pointwise error is computed and the maximum is reported.

This section validates the solver templates, not the LLM. Standard geometries are produced by deterministic geometry generators, and the FEniCS scripts are deterministic functions of the validated specifications. The LLM-facing parser and custom geometry path are evaluated separately in Section~\ref{sec:frontend_benchmarks}.

\subsection{Linear Elasticity}\label{sec:linear_elasticity}

For two-dimensional linear elasticity, we run a cantilever beam under end load and a plate with a circular hole under uniaxial tension. The cantilever uses six mesh levels and converges to the beam theory tip deflection. The plate-with-hole case uses a quarter-plate with symmetric boundary conditions. The plate is large enough relative to the hole to approximate the infinite-plate Kirsch solution.

Stress extraction at a stress concentration is sensitive to projection. A global $L^2$ projection of von Mises stress smears the peak by 10 to 15\%. The agent instead projects stress tensor components separately to a CG1 space and computes von Mises stress analytically at the critical point. With this extraction method, the plate-with-hole case reaches the Kirsch stress concentration factor of 3 at moderate mesh density and remains stable under further refinement.

The Lam\'{e} thick-walled cylinder under internal pressure is a third linear elastic check. The radial von Mises stress is compared against the closed-form Lam\'{e} solution. The error in maximum stress drops from double-digit percentages on the coarsest mesh to below 1\% on refined meshes.

For three-dimensional linear elasticity, we run a cantilever beam with quadratic tetrahedral elements. Tip deflection is compared against Timoshenko beam theory at five mesh densities. Stresses are evaluated at three interior cross-sections, $x = 20$, 50, and 100 mm from the clamp, avoiding the clamp singularity. All three stations reach sub-percent agreement with their Euler--Bernoulli reference stresses.

\begin{figure*}[h!]
\centering
\includegraphics[width=0.85\textwidth]{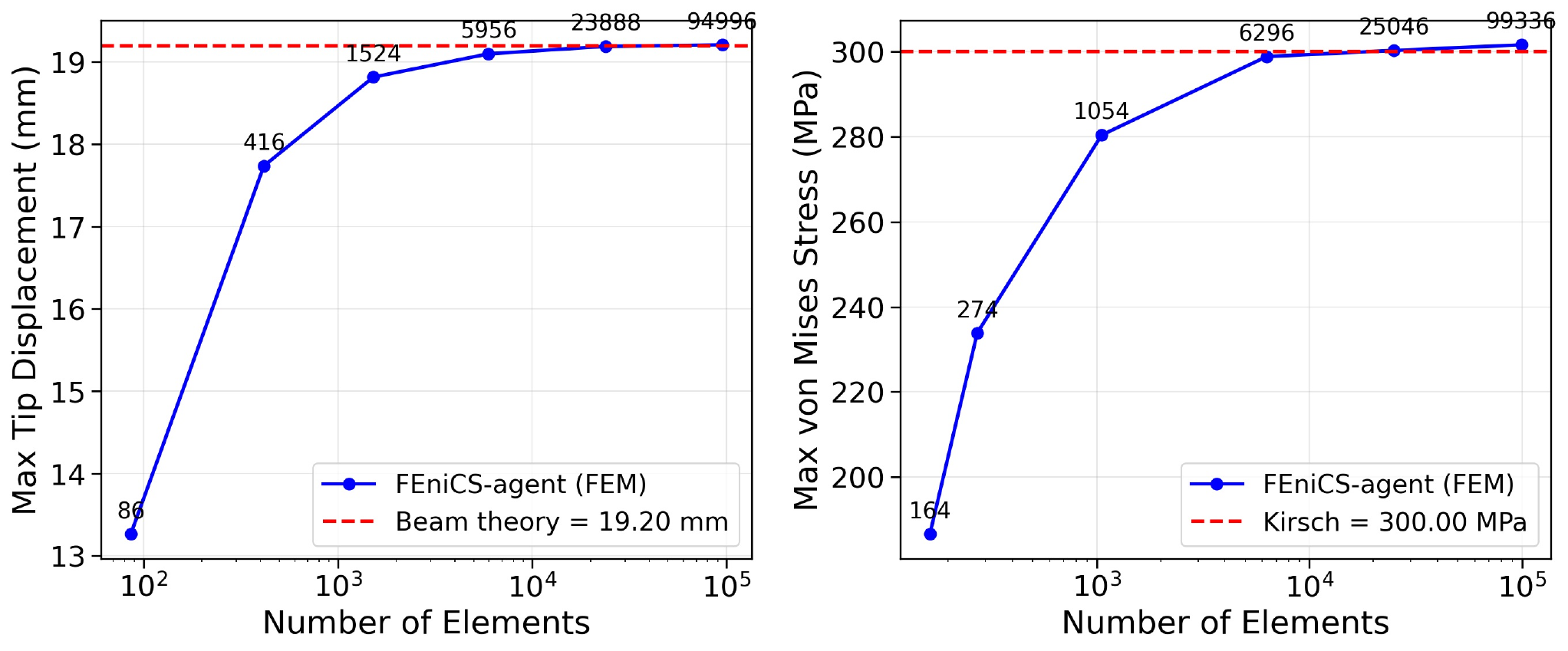}
\caption{Mesh convergence for two-dimensional linear elasticity. Left: cantilever tip deflection rises from 13.3 mm at 86 elements to 19.21 mm at about 95,000 elements, matching the beam theory value of 19.20 mm. Right: plate-with-hole maximum von Mises stress rises from 188 MPa at 164 elements to 300 MPa at about six thousand elements and beyond, matching the Kirsch stress concentration factor of 3 for an applied stress of 100 MPa.}
\label{fig:linear_2d}
\end{figure*}

\begin{figure*}[h!]
\centering
\includegraphics[width=0.85\textwidth]{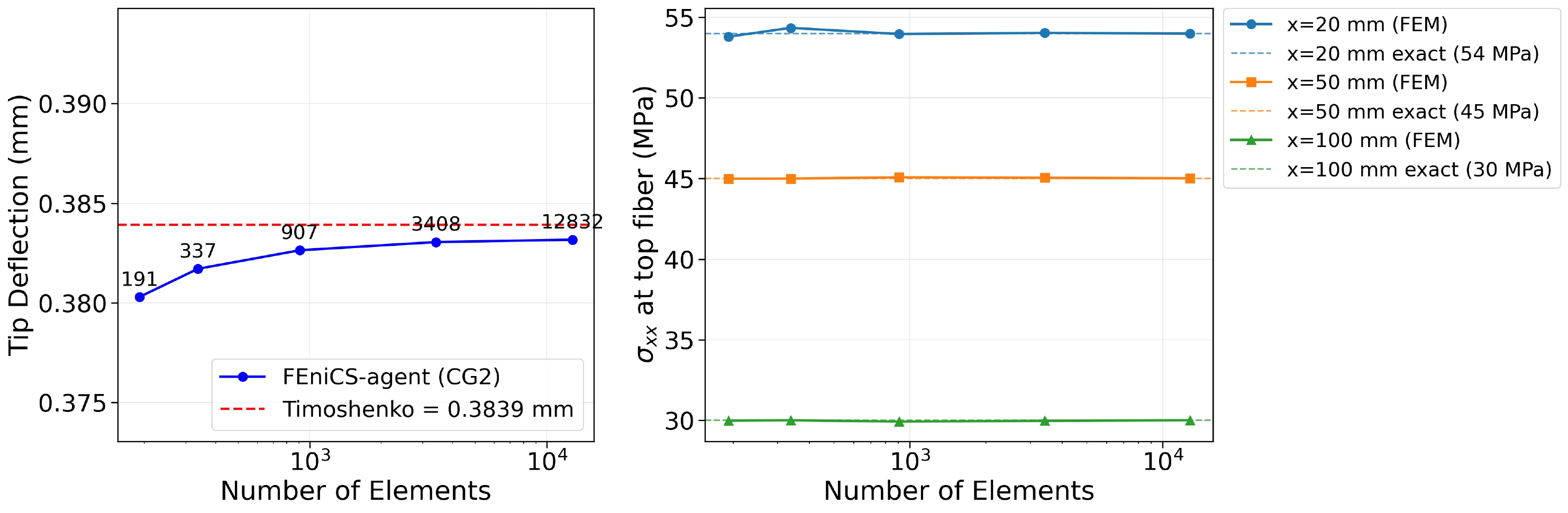}
\caption{Three-dimensional cantilever beam with quadratic tetrahedral elements. Left: tip deflection rises from 0.3803 mm to 0.3833 mm across five mesh densities, against the Timoshenko reference of 0.3839 mm. Right: bending stress at three interior cross-sections ($x = 20$, 50, and 100 mm), each compared against the corresponding Euler--Bernoulli value of 54, 45, and 30 MPa. All stations track their references within sub-percent error.}
\label{fig:linear_3d}
\end{figure*}

\begin{figure*}[h!]
\centering
\includegraphics[width=0.85\textwidth]{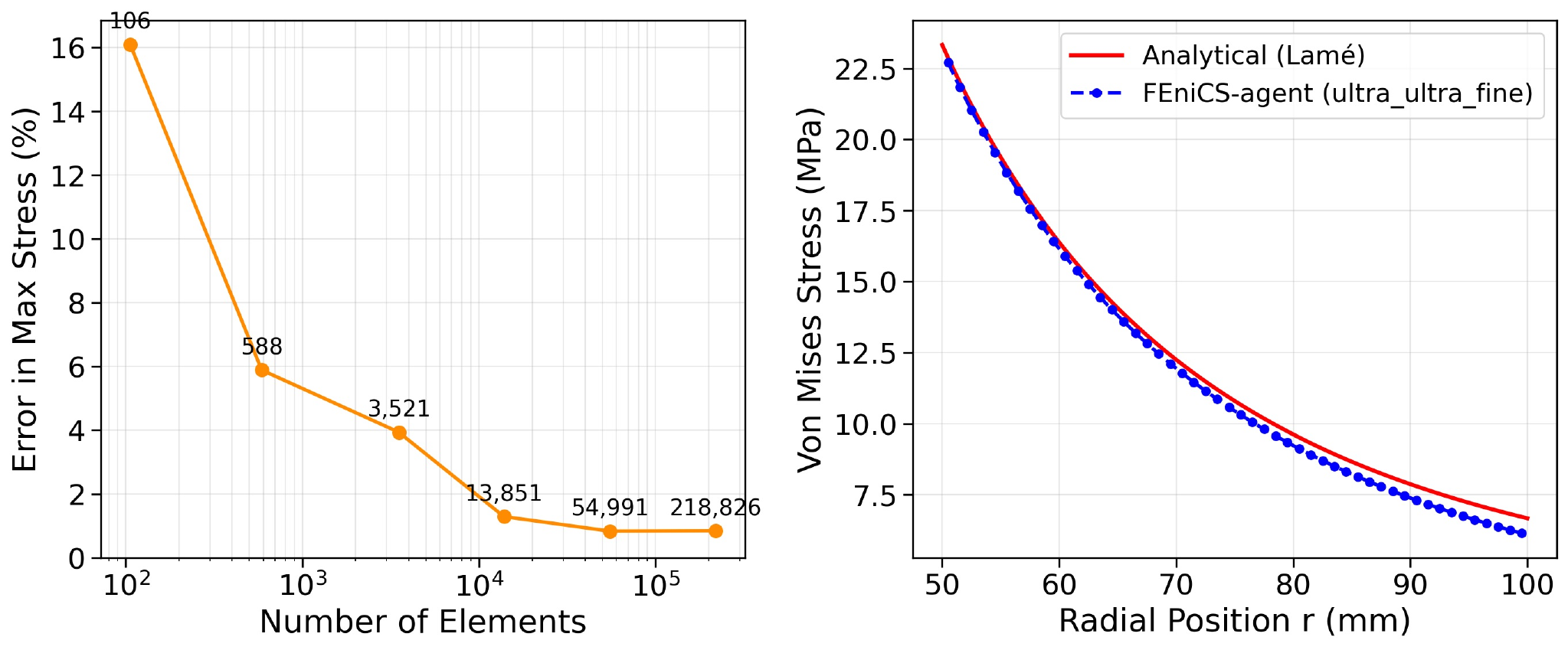}
\caption{Lam\'{e} cylinder verification. Left: error in maximum von Mises stress against the analytical Lam\'{e} value, dropping from 16\% at 106 elements to below 1\% at meshes finer than about 14,000 elements. Right: stress distribution along the radius at the finest mesh, overlaid with the closed-form Lam\'{e} solution.}
\label{fig:lame}
\end{figure*}

Figure~\ref{fig:linear_2d} shows convergence for the two-dimensional cantilever and plate-with-hole cases. Figure~\ref{fig:linear_3d} shows the three-dimensional cantilever convergence. Figure~\ref{fig:lame} shows the Lam\'{e} cylinder convergence and analytical stress profile.

\subsection{Hyperelasticity}\label{sec:hyperelasticity}

Cook's membrane is used as a nonlinear bending benchmark. We run a Neo-Hookean material with the Andelfinger and Ramm parameters and compare the upper-right tip displacement against the published reference of 23.91 mm. Linear triangles with $\nu = 1/3$ are tested across six mesh levels. The tip displacement converges from 21.82 mm on the coarsest mesh to within 0.21\% of the reference at high refinement.

\begin{figure*}[h!]
\centering
\includegraphics[width=0.85\textwidth]{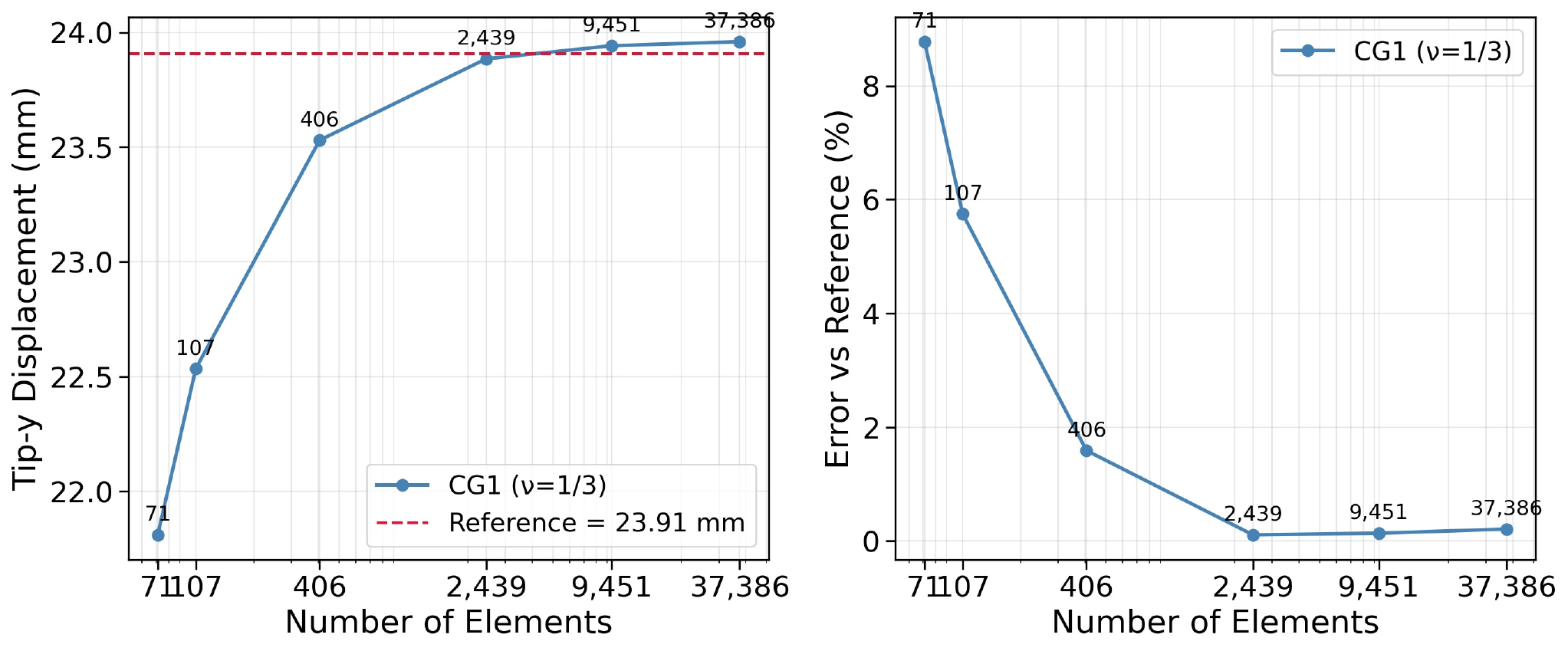}
\caption{Cook's membrane convergence with Neo-Hookean material and Poisson ratio $\nu = 1/3$. Left: tip-y displacement against the 23.91 mm reference of Andelfinger and Ramm across six mesh densities from 71 to 37,386 linear triangles. Right: percentage error against the reference, dropping from 8.79\% on the coarsest mesh to 0.21\% on the finest.}
\label{fig:cooks}
\end{figure*}

The three-dimensional rubber block under uniaxial compression provides a second hyperelastic check. With frictionless boundary conditions on the loaded face, the compressible Neo-Hookean response has a closed-form engineering-stress expression. This allows comparison at every load step. The relative error stays below $5 \times 10^{-5}$\% across the full compression range.

\begin{figure*}[h!]
\centering
\includegraphics[width=0.85\textwidth]{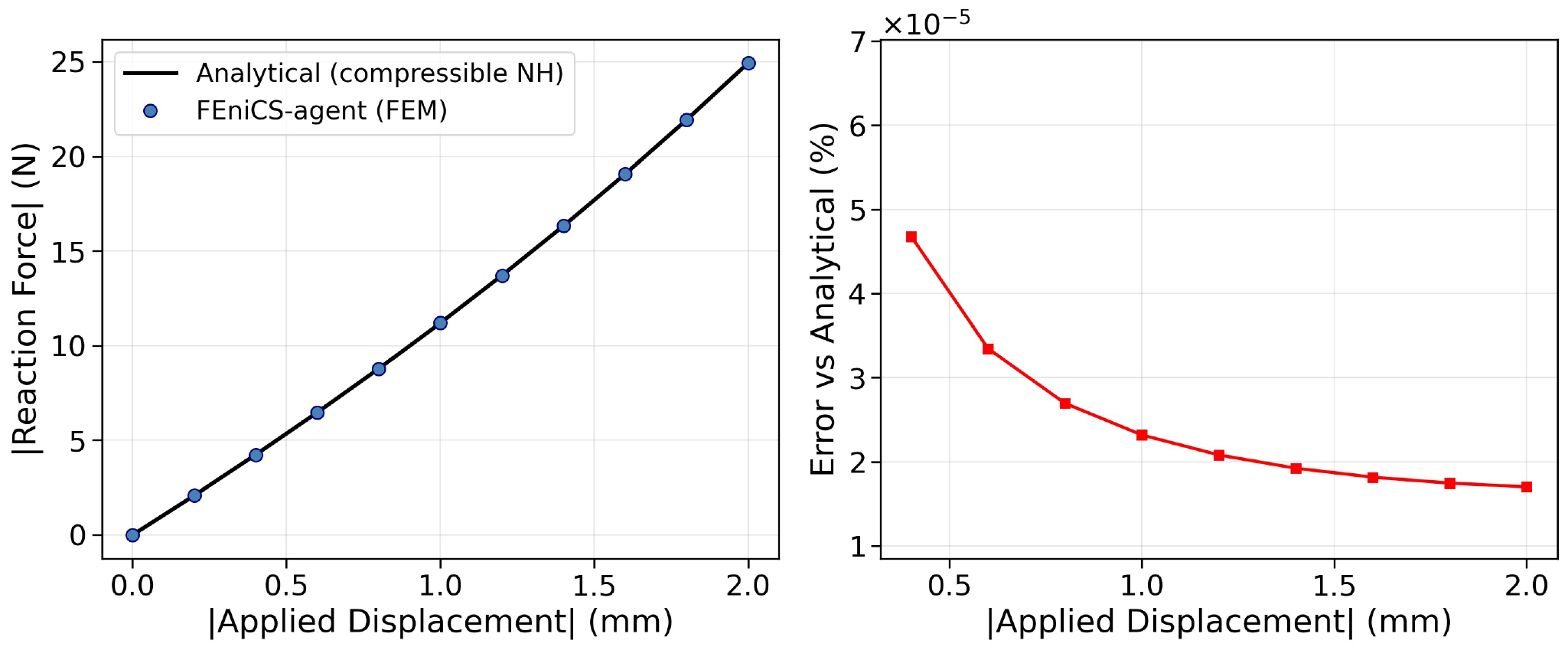}
\caption{Three-dimensional rubber block compression with frictionless boundary conditions. Left: FEM reaction force overlaid on the analytical compressible Neo-Hookean response across the full compression range from 0 to 2 mm. Right: percentage error against the analytical curve, which stays below $5 \times 10^{-5}$\% at every load step.}
\label{fig:rubber}
\end{figure*}

\subsection{Elastoplasticity}\label{sec:elastoplasticity}

A three-dimensional bar under uniaxial tension with linear isotropic hardening provides the elastoplastic benchmark. The bar is loaded through forty steps spanning elastic and plastic response. The elastic branch matches the analytical curve to within $10^{-4}$\%. Yield is detected at 250 MPa and 0.0595 mm displacement. The post-yield response approaches the analytical hardening line. The error spikes at the load step crossing the yield surface because the transition is resolved with finite increments, then settles to about 5\% at full load.

\begin{figure*}[h!]
\centering
\includegraphics[width=0.85\textwidth]{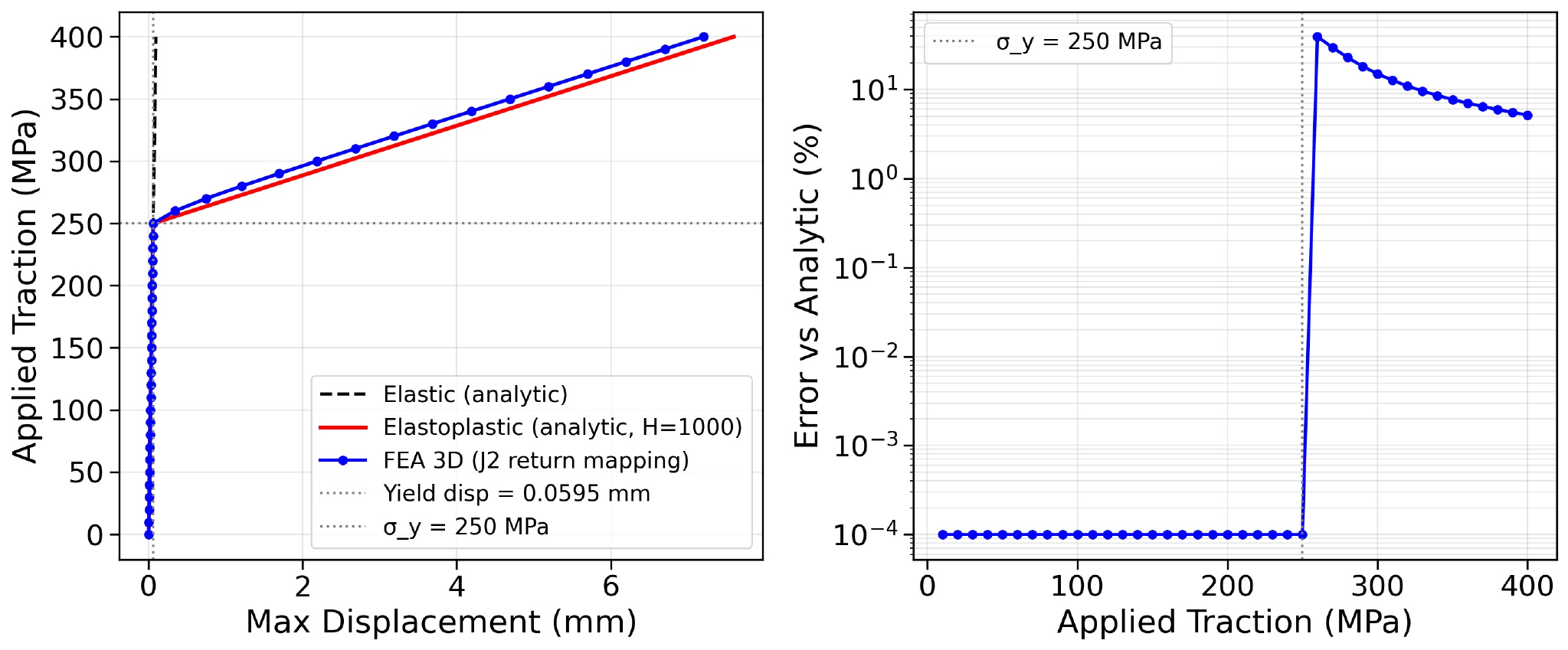}
\caption{Three-dimensional bar in uniaxial tension with linear isotropic hardening ($H = 1000$ MPa). Left: applied traction against maximum displacement, with the analytical elastic curve and hardening branch overlaid. Yield onset is detected at 250 MPa and 0.0595 mm. Right: percentage error against the analytical curve on a log scale, sitting at $10^{-4}$\% in the elastic regime, jumping at yield onset due to load-step discretization, and settling to about 5\% at full load.}
\label{fig:plastic}
\end{figure*}

\subsection{Thermo-Mechanical Coupling}\label{sec:thermo_mechanical}

\begin{figure}[h!]
\centering
\includegraphics[width=\linewidth]{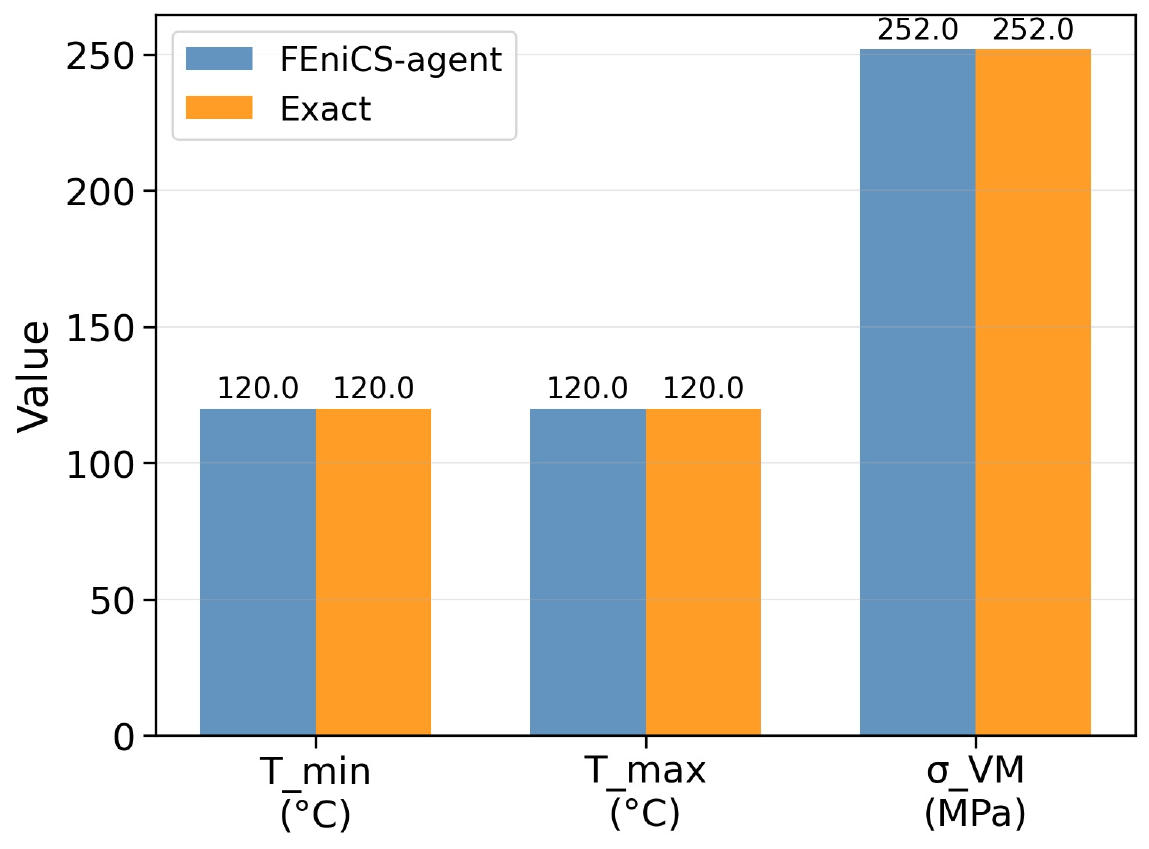}
\caption{Three-dimensional thermo-mechanical beam with both ends fixed under uniform heating. The agent reproduces \(T_{\min} \approx T_{\max} \approx 120~^\circ\)C and the analytical restrained thermal stress \(E\alpha\Delta T = 252\) MPa to within \(10^{-3}\%\).}
\label{fig:thermal}
\end{figure}

A three-dimensional steel beam fixed at both ends under uniform heating provides an exact reference for restrained thermal stress. Both thermal boundary faces are set to 120~\(^\circ\)C, with a reference temperature of 20~\(^\circ\)C, so \(\Delta T = 100~^\circ\)C. The corresponding restrained uniaxial thermal stress is \(\sigma_{xx} = -E\alpha\Delta T\), and the von Mises reference is \(E\alpha\Delta T = 252\) MPa for the chosen parameters. The computed temperature field is uniform to numerical precision, with \(T_{\min} \approx T_{\max} \approx 120~^\circ\)C. The maximum von Mises stress matches the analytical reference to within \(10^{-3}\%\). Figure~\ref{fig:thermal} shows the comparison.

\subsection{Phase-Field Fracture}\label{sec:phase_field}

The single edge notched tension test is used for phase-field fracture. We compare against the reference of Miehe et al. for two-dimensional and three-dimensional cases. The two-dimensional case shows a linear elastic rise to a peak reaction force of about 596 N/mm, followed by a sharp drop as the damage field reaches unity. Fracture onset matches the Miehe reference of $5.60 \times 10^{-3}$ mm to within roughly 2\%.

The three-dimensional case is set up as a thin slab with plane-strain boundary conditions on the front and back faces. This lets the template exercise a three-dimensional mesh and vector displacement field while keeping the element count manageable. The 3D slab gives a raw peak reaction force of about 60.8 N; dividing by the 0.1 mm slab thickness gives about 608 N/mm, consistent with the 2D per-thickness scale. The crack path follows the expected horizontal trajectory from the notch tip in both cases.

\begin{figure*}[h!]
\centering
\includegraphics[width=0.85\textwidth]{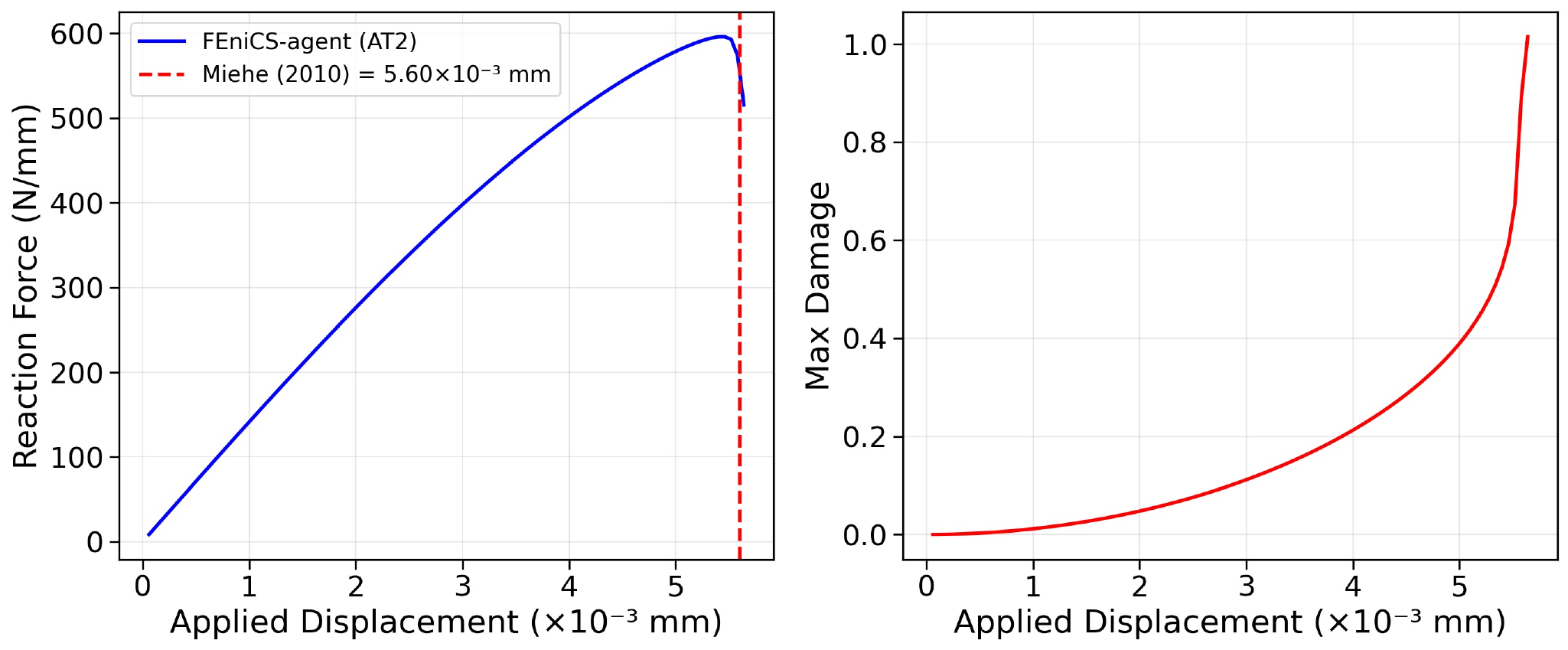}
\caption{Two-dimensional SENT phase-field fracture. Left: reaction force per unit thickness against applied displacement, showing linear elastic rise, peak load near 596 N/mm, and sharp drop at fracture. Right: maximum damage against applied displacement. Fracture onset matches the Miehe reference of $5.60 \times 10^{-3}$ mm to within roughly 2\%.}
\label{fig:sent_2d}
\end{figure*}

\begin{figure*}[h!]
\centering
\includegraphics[width=0.85\textwidth]{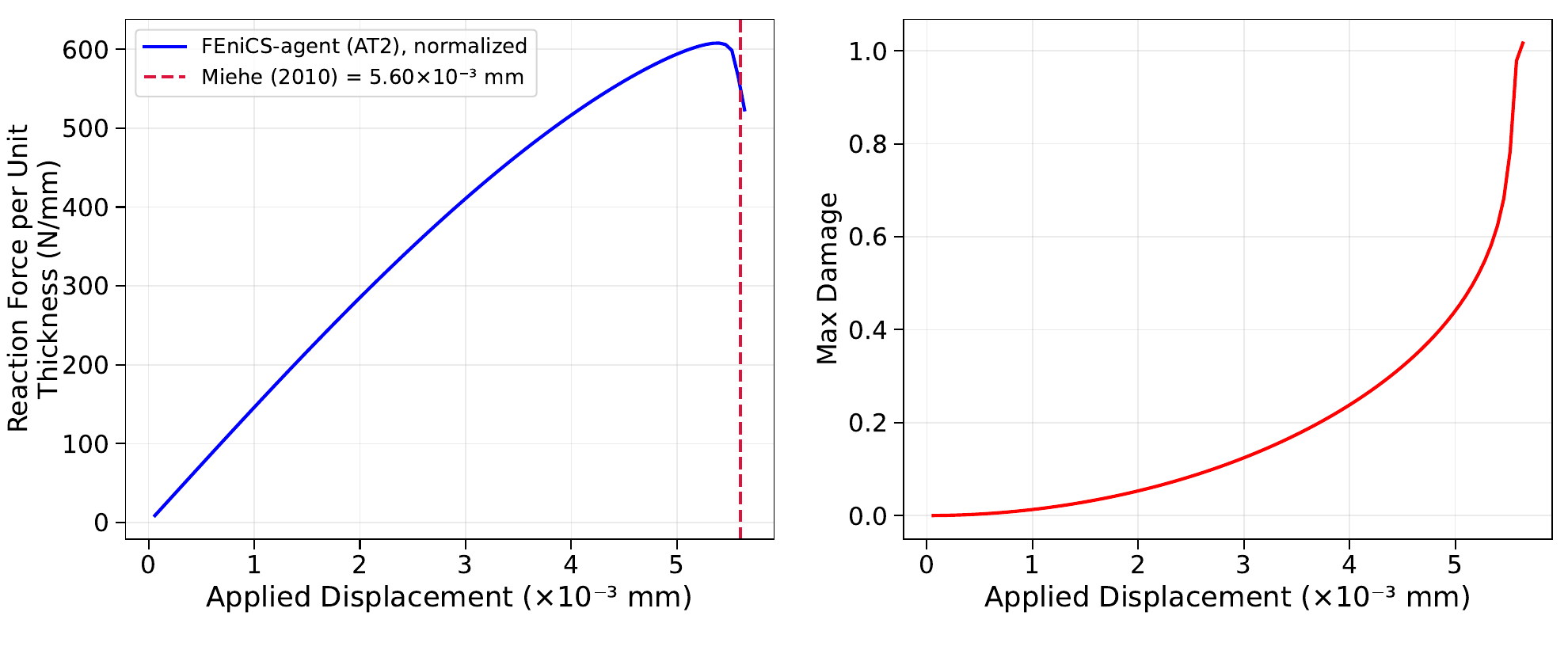}
\caption{Three-dimensional SENT phase-field fracture. A thin slab with plane-strain boundary conditions gives a raw peak force near 60.8 N, corresponding to about 608 N/mm after division by the 0.1 mm thickness, with the same fracture-onset displacement.}
\label{fig:sent_3d}
\end{figure*}

\begin{figure}[h!]
\centering
\includegraphics[width=\linewidth]{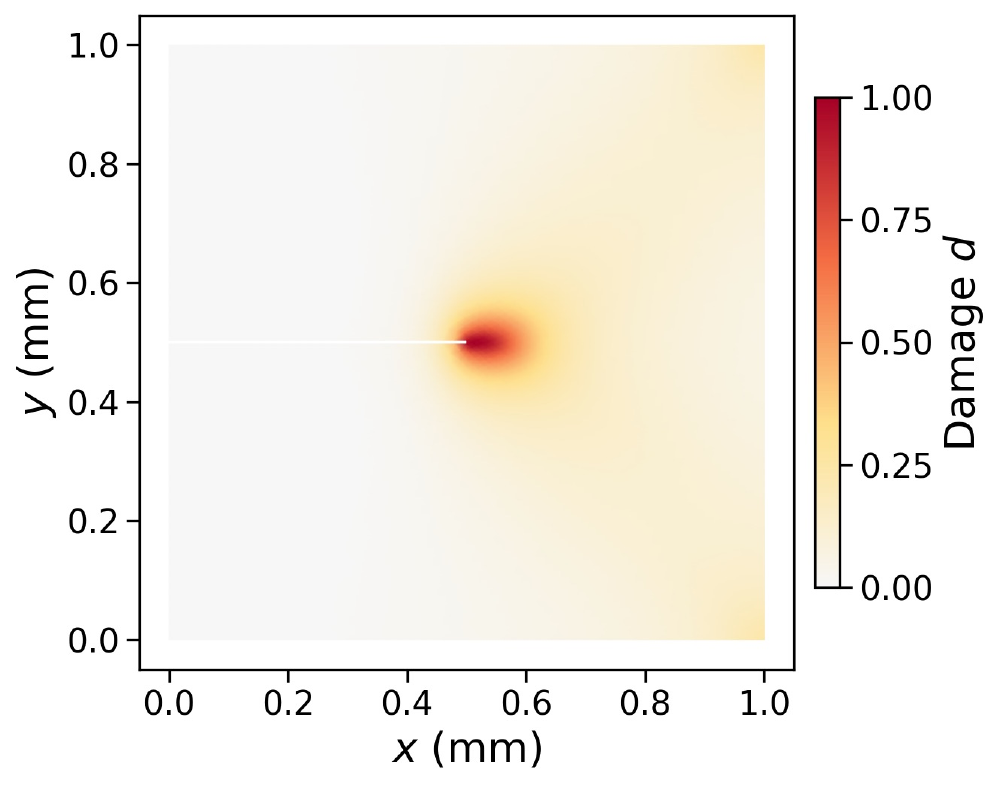}
\caption{Damage field in the two-dimensional SENT specimen at full fracture. The crack is visible as a localized concentration of $d \to 1$ at the notch tip.}
\label{fig:damage_2d}
\end{figure}

\begin{figure*}[h!]
\centering
\includegraphics[width=0.85\textwidth]{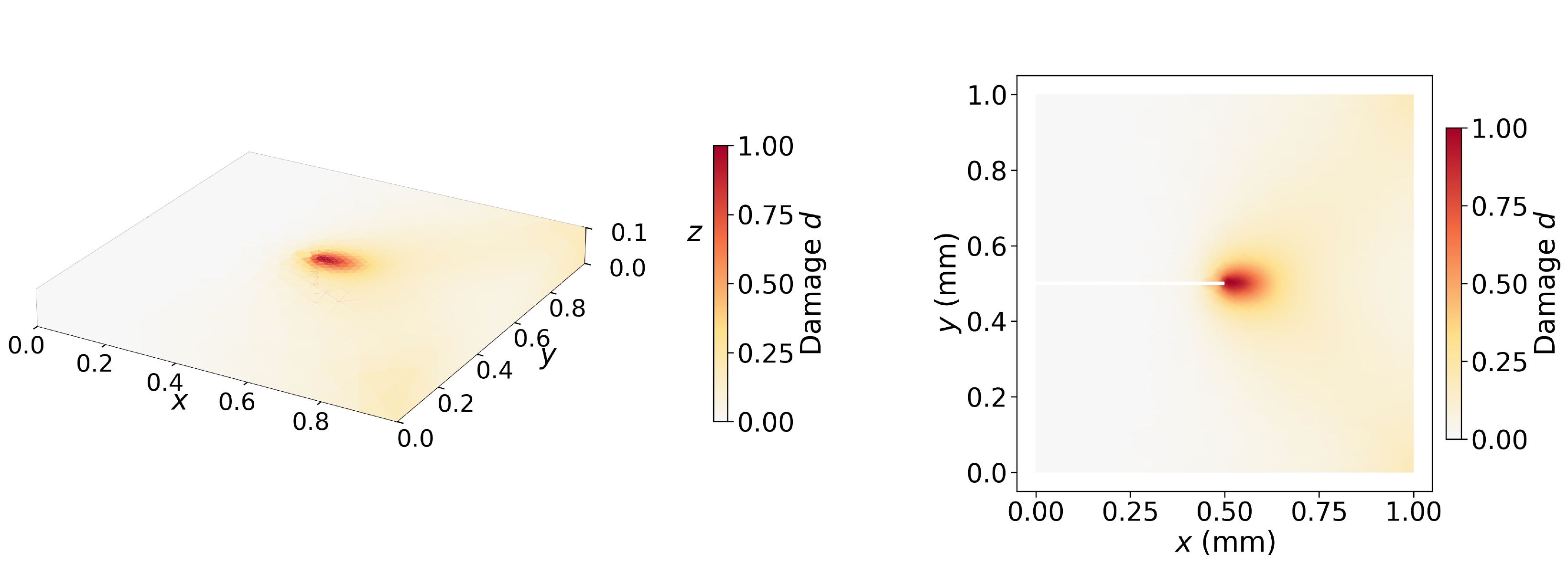}
\caption{Damage field on the three-dimensional SENT mesh, shown in perspective on the left and as a top-down projection on the right. The crack forms a localized region at the notch tip, consistent with the plane-strain setup.}
\label{fig:damage_3d}
\end{figure*}

Figures~\ref{fig:sent_2d} and~\ref{fig:sent_3d} show load-displacement curves and maximum damage evolution. Figures~\ref{fig:damage_2d} and~\ref{fig:damage_3d} show the localized damage field.

\subsection{Validation Summary}\label{sec:validation_summary}

Table~\ref{tab:validation} collects the deterministic mechanics validation results. All ten cases reach quantitative agreement with analytical or published references. Smooth linear and nonlinear cases reach sub-percent agreement on adequate meshes. The harder nonlinear cases, elastoplasticity past yield and phase-field fracture, reach the 2--5\% range. The plasticity error is dominated by load-step discretization at the elastic-plastic transition. The phase-field error is dominated by length-scale resolution against the Miehe reference. These results show that the deterministic template layer works across the five supported physics classes. They do not, by themselves, measure the LLM interface; that is the purpose of the next section.

\begin{table}[h!]
\centering
\caption{Validation summary across deterministic mechanics benchmarks. Errors are measured as relative percentage errors against analytical solutions where available and against published numerical benchmarks otherwise.}
\label{tab:validation}
\small
\begin{tabular}{llll}
\toprule
Problem & Dim & Reference & Error \\
\midrule
Cantilever beam & 2D & Beam theory & $<$0.1\% \\
Plate with hole & 2D & Kirsch SCF & $<$0.5\% \\
Lam\'{e} cylinder & 2D & Analytical & $<$1\% \\
Cantilever beam & 3D & Timoshenko & $<$0.2\% \\
Cook's membrane & 2D & Andelfinger \& Ramm & 0.21\% \\
Rubber block & 3D & Neo-Hookean & $<10^{-4}$\% \\
Bar plasticity & 3D & Analytical hardening & $\sim$5\% \\
Heated beam & 3D & $E\alpha\Delta T$ & $<10^{-3}$\% \\
SENT fracture & 2D & Miehe et al. & $\sim$2\% \\
SENT fracture & 3D & Miehe et al. & $\sim$2\% \\
\bottomrule
\end{tabular}
\end{table}

\subsection{Heuristic Local-Refinement Study}\label{sec:local_refinement}

The meshing pipeline uses heuristic local-refinement rules rather than a posteriori error estimation. To check whether these rules are actually enforced and whether they improve accuracy per cost, we ran a three-case study with the same three mesh strategies in each case: a uniform coarse mesh, a heuristic locally refined mesh, and a uniform fine mesh. The cases were a two-dimensional plate with a hole, a two-dimensional SENT phase-field fracture case, and a three-dimensional notched box. The three-dimensional notched-box case is a rectangular solid with a localized side notch, used here as a simple 3D test of the notch-targeted local-refinement rule in the meshing backend. These cases were chosen because each has a clear local refinement target already encoded in the meshing backend.

This study is separate from the final validation summary in Table~\ref{tab:validation}. The plate-with-hole error uses the analytical 300 MPa Kirsch reference, whereas the SENT and notched-box errors use the best available uniform-fine solution as an internal reference. The errors in this study therefore should not be compared directly with the final validation errors in Table~\ref{tab:validation}.

We first checked whether the requested local refinement was actually applied in the generated mesh. In all three cases, the observed local mesh size near the target region matched the requested heuristic rule to within a small tolerance, so the implemented local-refinement rules were enforced quantitatively. We then compared the relevant quantity of interest across the three mesh strategies. For the plate-with-hole case, the locally refined mesh reduced the stress error from 37.13\% to 22.91\%. For the SENT phase-field case, using peak reaction force as the primary comparison metric, the locally refined mesh reduced the error from 4.65\% to 0.44\%. For the three-dimensional notched-box case, the locally refined mesh reduced the displacement error from 9.43\% to 1.91\%. For the three-dimensional notched-box case, the uniform fine mesh is used as the internal reference for this study, so the reported errors measure deviation from that fine-mesh result rather than deviation from an analytical solution.

\begin{figure}[h!]
\centering
\includegraphics[width=\linewidth]{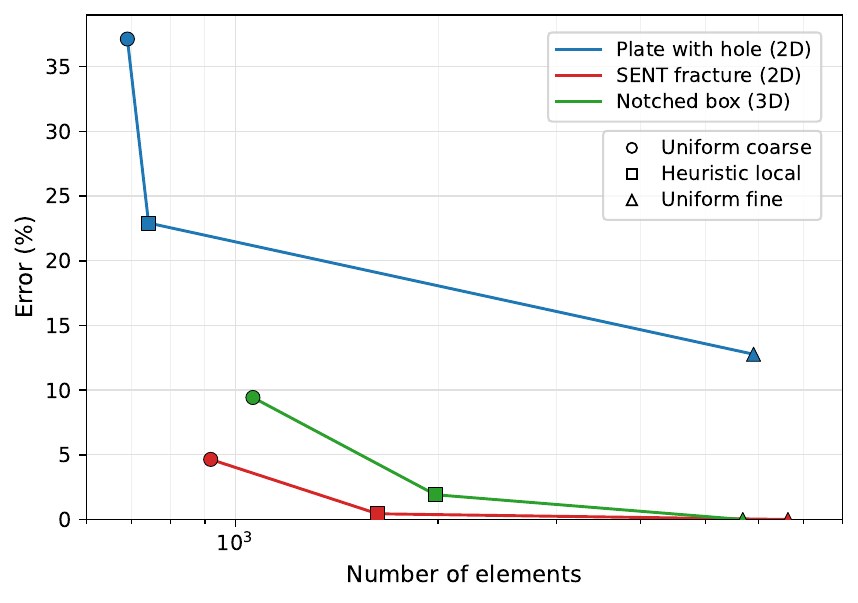}
\caption{Heuristic local-refinement study across three deterministic benchmark cases. This study is separate from the final validation summary in Table~\ref{tab:validation}. Errors are reported relative to the case reference used in the study. The locally refined meshes enforce the intended refinement rules and improve the primary quantity of interest relative to the coarse uniform meshes while remaining much smaller than the uniform fine meshes.}
\label{fig:local_refinement_study}
\end{figure}

These results support a narrow claim. The current backend does enforce its intended hole and notch-based local refinement rules, and those rules can improve accuracy in representative two and three-dimensional cases without requiring the much larger uniform fine meshes. At the same time, the refinement logic remains heuristic. In particular, the phase-field refinement is not derived automatically from the fracture length scale $\ell$; it is only activated when a notch or crack-band refinement region is requested explicitly with a suitable local mesh size. Figure~\ref{fig:local_refinement_study} summarizes the study.

\section{LLM-Facing Front-End Evaluation}\label{sec:frontend_benchmarks}

The mechanics benchmarks in Section~\ref{sec:mechanics_validation} test the deterministic backend. They do not measure the natural-language interface. We therefore evaluate the two places where the LLM affects the pipeline: structured specification generation and custom geometry generation. These benchmarks use the same parser, validator, retry logic, and geometry checks used by the full system.

\subsection{Parser Robustness and Repair}\label{sec:parser_robustness}

The parser benchmark contains 15 prompts: 10 accept cases, 2 ambiguous cases, and 3 invalid cases. It is not a classification-only accept/reject test. It measures whether the front end can turn a prompt into a valid structured specification after validation and, when needed, repair. A first-pass parse is counted as valid if it satisfies the JSON schema and passes the validator without retry. A repaired parse is counted when the first attempt fails validation but a later retry produces a valid specification. The invalid cases in this benchmark are not counted as successful clean rejections; they are cases where validation feedback repaired an initially invalid or inconsistent prompt into a physically admissible specification. The benchmark therefore measures repair into a valid structured specification, not conservative refusal behavior.

First-pass valid parses were obtained for 9 of 15 cases (60.0\%). The remaining 6 cases (40.0\%) failed validation on the first attempt but were repaired after retry. The final valid parse rate was therefore 15 of 15 cases (100.0\%). On labeled accept cases, problem-class accuracy was 100.0\%, the wrong problem-class rate was 0.0\%, and field-extraction accuracy was 97.1\%. The only field-extraction miss was a thermo-mechanical prompt where the parser omitted \texttt{temperature} from the requested output fields.

These results support a specific claim. The parser is a repair-oriented structured-specification front end. It is not a strict clean-rejection system. About two fifths of the benchmark cases needed correction, but all cases eventually produced a valid specification. The perfect problem-class accuracy on this small benchmark is useful, but it should not be treated as the whole result. The main result is that validation and targeted retry convert first-pass specification failures into valid structured inputs.

\begin{table}[h!]
\centering
\caption{Parser robustness benchmark. The benchmark measures structured-specification generation under validation and retry, not only first-pass classification.}
\label{tab:parser_benchmark}
\small
\renewcommand{\arraystretch}{1.15}
\begin{tabular}{lr}
\toprule
Metric & Value \\
\midrule
Total prompts & 15 \\
Accept / ambiguous / invalid & 10 / 2 / 3 \\
First-pass valid parses & 9 / 15 (60.0\%) \\
First-pass validation failures & 6 / 15 (40.0\%) \\
Repaired after retry & 6 / 15 (40.0\%) \\
Final valid parses & 15 / 15 (100.0\%) \\
Problem-class accuracy & 100.0\% \\
Wrong problem-class rate & 0.0\% \\
Field-extraction accuracy & 97.1\% \\
\bottomrule
\end{tabular}
\end{table}

The problem-class confusion matrix is effectively diagonal on the 10 labeled accept cases in the five physics classes used in the paper. Because the benchmark is about structured specifications rather than classification alone, and because the scalar accuracy summary already captures the class result, we treat the full matrix as supplementary material.

\subsection{Custom Geometry Generation}\label{sec:custom_geometry}

The custom geometry benchmark evaluates the real LLM-to-Gmsh path. This path is harder than parsing because the model must generate executable geometry code, identify surfaces after boolean operations, and assign physical boundary tags that the solver can use. Each case is counted as a first-pass success only if the first attempt passes syntax checks, security checks, geometry validation, mesh generation, and boundary-tag validation. Retry recovery is counted only when an initial failure is fixed by the feedback loop and the final geometry passes the same checks.

Across 10 custom-geometry prompts, first-pass generation and meshing succeeded for 9 cases (90.0\%). No additional cases were recovered by retry, so retry recovery was 0 of 10 cases (0.0\%). The final success rate was therefore 9 of 10 cases (90.0\%), with 1 unrecovered failure (10.0\%). Attempt-level failures were all classified as invalid geometry: \texttt{invalid\_geometry} occurred 3 times, while \texttt{invalid\_boundary\_tags}, syntax, and other failures did not occur. The one final unrecovered failure was also \texttt{invalid\_geometry}. It occurred in a harder obround-slot case that remained unrecovered after three attempts. All three attempts failed the geometry validation checks, suggesting that the current prompt and validation design cannot yet synthesize this class of compound-slot geometry reliably from primitives alone.

These results are specific to this benchmark. The current constrained prompt and validation design is effective on this 10-case set, but it is not evidence of broad reliability for arbitrary CAD-like parts. In the current system, LLM-generated Gmsh geometry is a measured and useful path for some non-catalog shapes, but it is still not a general CAD replacement.

\begin{table}[h!]
\centering
\caption{Custom geometry-generation benchmark for the real LLM-to-Gmsh path. Final success requires valid code, valid geometry, mesh generation, and required boundary tags.}
\label{tab:geometry_benchmark}
\small
\renewcommand{\arraystretch}{1.15}
\begin{tabular}{lr}
\toprule
Metric & Value \\
\midrule
Total custom geometry cases & 10 \\
First-pass successes & $9/10$ (90.0\%) \\
Retry recoveries & $0/10$ (0.0\%) \\
Final successes & $9/10$ (90.0\%) \\
Final failures & $1/10$ (10.0\%) \\
\midrule
Attempt-level \texttt{invalid\_geometry} & 3 \\
Attempt-level \texttt{invalid\_boundary\_tags} & 0 \\
Attempt-level \texttt{syntax} & 0 \\
Attempt-level \texttt{other} & 0 \\
\midrule
Final \texttt{invalid\_geometry} & 1 \\
Final \texttt{invalid\_boundary\_tags} & 0 \\
Final \texttt{mesh\_failure} & 0 \\
Final \texttt{other} & 0 \\
\bottomrule
\end{tabular}
\end{table}

\section{Parametric Studies and Optimization}\label{sec:parametric_studies}

\subsection{Parametric Study Capability}\label{sec:parametric_capability}

To demonstrate parametric mode, we run a 25-configuration study of a plate with a circular hole. Plate thickness varies from 1 to 5 mm, and hole radius varies from 10 to 50 mm. The agent expands the parameter ranges from one natural-language input, runs all configurations in parallel, and aggregates the results without manual intervention. Total wall-clock time is under three minutes on a workstation with eight worker processes.

\begin{figure}[h!]
\centering
\includegraphics[width=\linewidth]{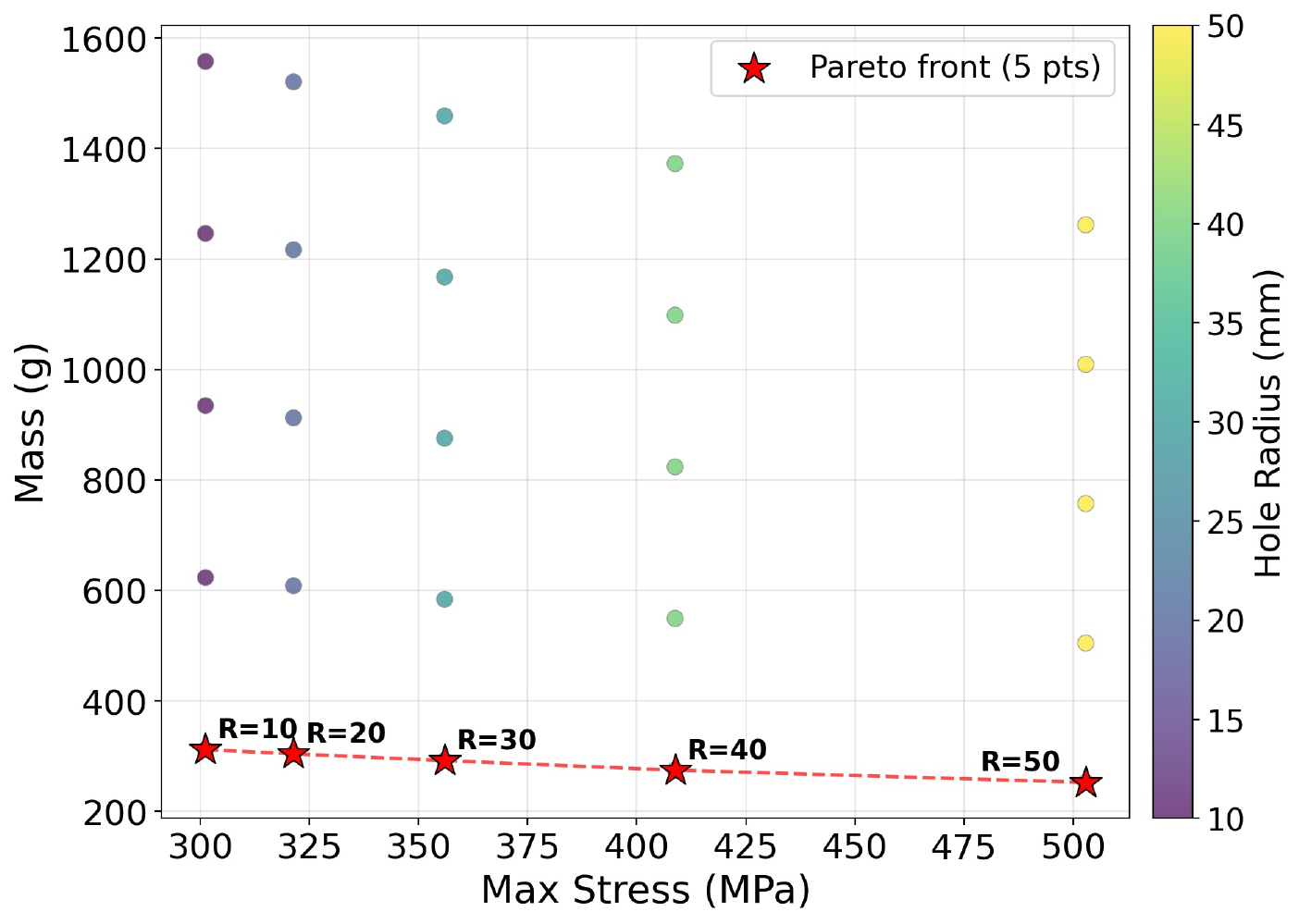}
\caption{Pareto front for the plate-with-hole parametric study. Five Pareto-optimal designs ($R = 10$ to 50 mm) form a mass-stress trade-off curve at the smallest thickness, with mass between roughly 250 and 310 g and maximum stress between roughly 300 and 510 MPa. Dominated designs are shown as circles where color indicates hole radius.}
\label{fig:pareto}
\end{figure}

\begin{figure}[h!]
\centering
\includegraphics[width=\linewidth]{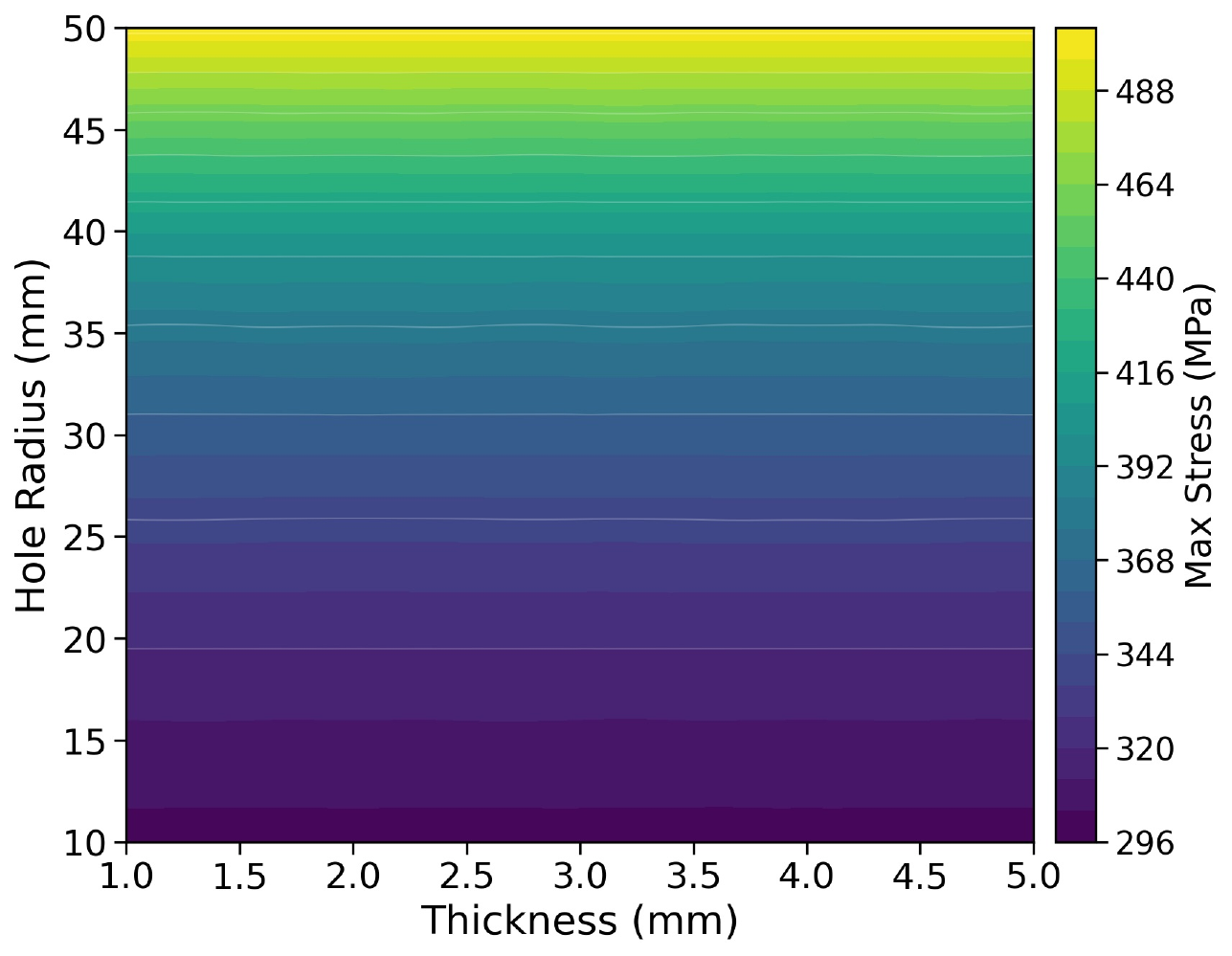}
\caption{Maximum von Mises stress contour over the thickness--hole-radius design space, ranging from about 300 MPa at small holes to about 500 MPa at the largest hole radius. The contour bands are horizontal because in-plane stress is thickness-independent under traction loading in plane stress. Thickness controls mass, which is the second objective in Figure~\ref{fig:pareto}.}
\label{fig:contour}
\end{figure}

The sensitivity analysis shows that hole radius drives stress (Pearson $r = 0.97$), while thickness controls mass. This is the expected behavior for a traction-loaded plate in plane stress, where the in-plane stress field is independent of thickness. The Pareto front in the mass-stress plane (Figure~\ref{fig:pareto}) identifies five Pareto-optimal designs, one at each hole radius, all at the lightest thickness. They form a trade-off from low mass and high stress at $R = 50$ mm to higher mass and lower stress at $R = 10$ mm. Figure~\ref{fig:contour} shows the maximum-stress contour over the design space. The horizontal bands reflect the thickness-independence of in-plane stress in this loading mode.

\subsection{Gradient-Free Design Optimization}\label{sec:optimization}

As a separate constrained optimization benchmark, we minimized the maximum von Mises stress in the plate-with-hole problem with the radius bounded to 5--35 mm. The v2 run used a bounded scalar search with explicit endpoint checks, and all evaluated physical radii remained inside the prescribed bounds. The best feasible evaluation occurred at the lower bound, \(R = 5.0\) mm, with maximum stress 189.46 MPa and maximum displacement 0.04811 mm. This result is reported as a separate optimization benchmark and is not compared to the 25-point parametric sweep above.

\begin{table}[h!]
\centering
\caption{Saved v2 constrained optimization benchmark for the plate-with-hole problem.}
\label{tab:optimization_result}
\small
\renewcommand{\arraystretch}{1.15}
\begin{tabular}{ll}
\toprule
Optimizer & Bounded scalar search + endpoint checks \\
Radius bounds & 5--35 mm \\
Objective & Minimize max von Mises stress \\
Constraint & Max stress $< 500$ MPa \\
Evaluations & 20 \\
Final hole radius (mm) & 5.0000 \\
Final max stress (MPa) & 189.46 \\
Final max displacement (mm) & 0.04811 \\
Optimum at bound & Yes, lower bound \\
All physical radii within bounds & Yes \\
\bottomrule
\end{tabular}
\end{table}

The black-box approach is general but expensive. It works with path-dependent operations such as J2 return mapping and with staggered phase-field solves because it does not differentiate through the solver. Adjoint methods such as those in dolfin-adjoint are better suited for differentiable forward problems, but they require special handling when the pipeline includes external geometry generation, non-smooth updates, or staggered history fields. The current optimizer is therefore best viewed as low-dimensional bounded gradient-free design optimization, not topology optimization.

\section{End-to-End Demonstration: 3D L-Bracket}\label{sec:lbracket}

The benchmark suite validates the deterministic templates on standard problems. The front-end benchmarks measure parser and custom geometry reliability across small prompt sets. We also include one successful end-to-end demonstration on a non-catalog geometry: a three-dimensional L-bracket with a fillet at the inner corner and a bolt hole in the horizontal arm. The case combines LLM-generated geometry, adaptive meshing at stress concentrations, elastoplastic material behavior, load stepping, and field post-processing.

This example should be read together with the geometry benchmark in Section~\ref{sec:frontend_benchmarks}. The L-bracket shows the full workflow on one successful non-catalog geometry. This case is intended as a workflow demonstration on a successful non-catalog geometry, not as a mesh-converged engineering assessment of the bracket.

\subsection{Natural-Language Input}\label{sec:lbracket_input}

The full prompt to the agent reads: ``\textit{Analyze a 3D steel L-bracket under shear loading. The bracket has a vertical arm 20 mm wide and 80 mm tall and a horizontal arm 80 mm long and 20 mm tall, both 10 mm thick. Add a 5 mm fillet at the inner corner and a through-hole of radius 4 mm centered at 60 mm along the horizontal arm. Material is structural steel with yield stress 250 MPa and hardening modulus 1000 MPa. Fix the bottom of the horizontal arm. Apply 50 MPa downward shear traction on the top face of the vertical arm. Use elastoplastic analysis with 40 load steps. Report stress distribution, plastic zone evolution, and force-displacement curve.}''

The parser produces a structured specification with custom geometry enabled, steel material properties, a fixed underside, a downward shear traction, and elastoplastic analysis with forty load steps. No fields in the specification were edited by hand.

\subsection{LLM Geometry Generation}\label{sec:lbracket_geometry}

The geometry stage routes the specification to the LLM-to-Gmsh path. The generated Gmsh OCC script creates two boxes for the arms, fuses them, subtracts a cylinder for the bolt hole, applies a fillet to the inner edge, and tags the loaded and fixed surfaces as physical groups. The code passes syntax and security checks, executes in the restricted execution namespace, and produces a geometry whose volume and surface tags pass validation. The model does not call a built-in L-bracket generator. It constructs the shape from primitives based on the prompt.

\subsection{Results}\label{sec:lbracket_results}

\begin{figure*}[h!]
\centering
\includegraphics[width=0.95\textwidth]{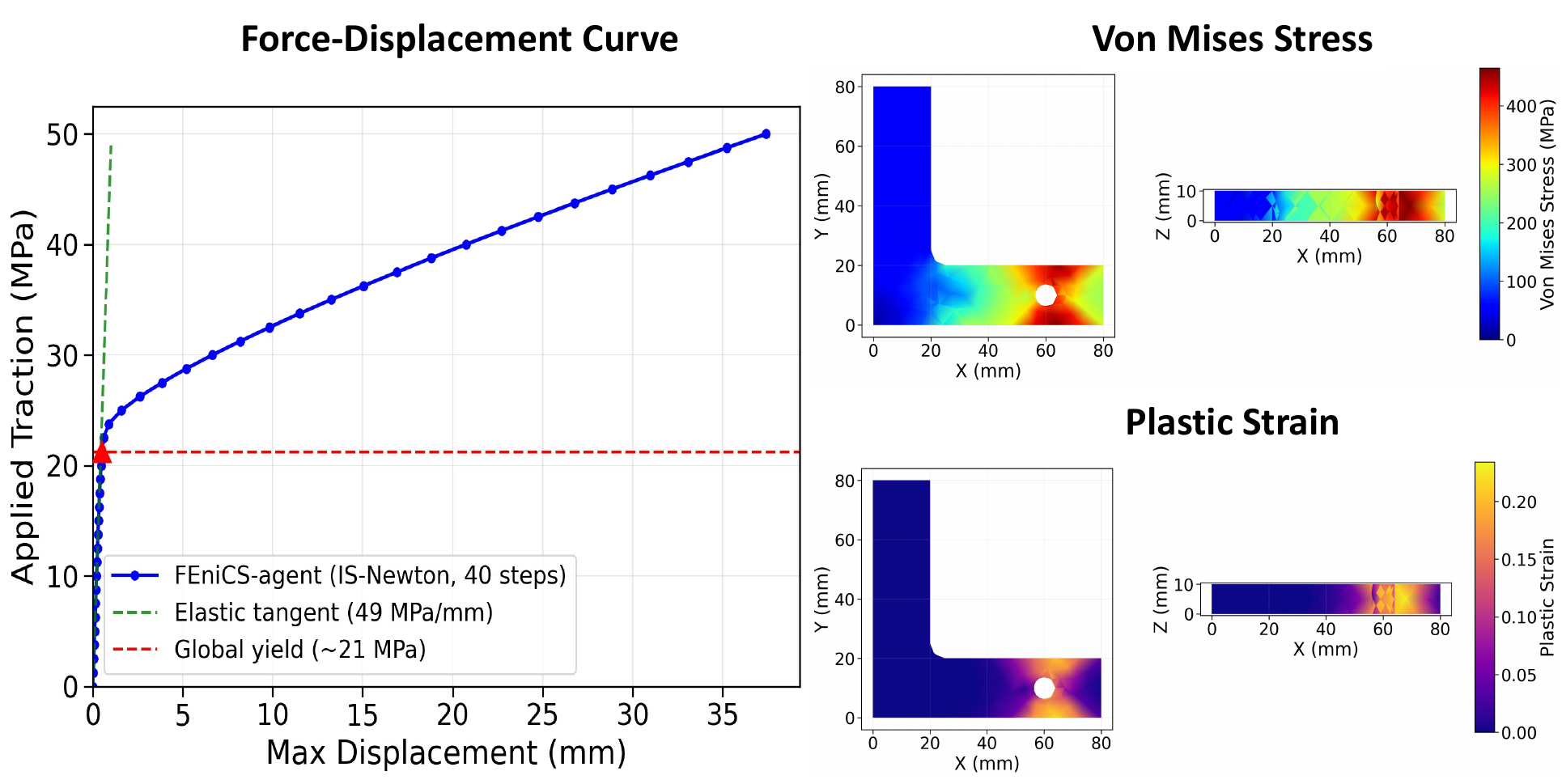}
\caption{End-to-end demonstration on a 3D L-bracket with elastoplasticity, generated from one natural-language prompt. Left: applied traction against maximum componentwise displacement from the load-step history, with the elastic tangent of 49 MPa/mm, global yield at about 21 MPa applied traction, and the post-yield branch reaching 50 MPa applied traction at about 37.4 mm maximum componentwise displacement. Top right: von Mises stress at peak load, with peak values above 400 MPa concentrated at the bolt hole and inner fillet. Bottom right: accumulated plastic strain at peak load, with peak values above 0.20 at the bolt hole.}
\label{fig:lbracket}
\end{figure*}

The full run takes 87.5 seconds, including parsing, geometry generation, meshing, template assembly, and forty elastoplastic load steps. The computed response shows an elastic regime followed by stress and plastic-strain localization near the bolt hole and inner fillet. As load increases, stress redistributes and the computed accumulated plastic-strain field develops around both stress concentration features. In the plotted load-history curve, displacement denotes the maximum componentwise displacement from the saved load-step history, not a tip-displacement probe. Figure~\ref{fig:lbracket} shows the load-history response using the saved stepwise maximum componentwise displacement from the load-step history, together with the von Mises stress and accumulated plastic strain fields at peak load.

\section{Computational Performance}\label{sec:performance}

\subsection{Timing Breakdown}\label{sec:timing}

\begin{figure}[h!]
\centering
\includegraphics[width=\linewidth]{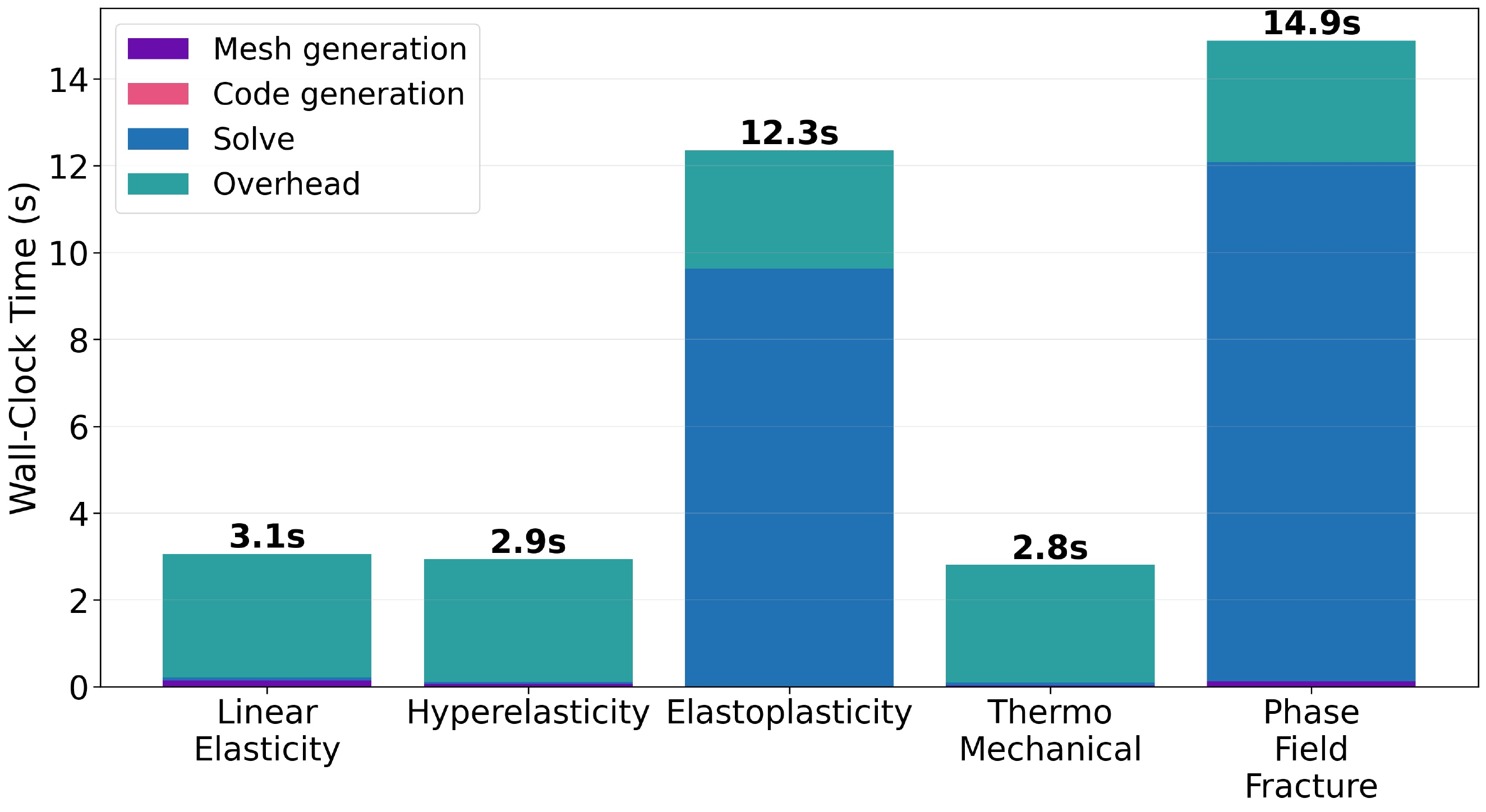}
\caption{Wall-clock time breakdown for one representative case from each physics class. Linear elasticity, hyperelasticity, and thermo-mechanical coupling complete in roughly 3 seconds because their benchmark solves are sub-second. Elastoplasticity (12.3 s, including 9.6 s for return mapping over twenty load steps) and phase-field fracture (14.9 s, including 12 s for the staggered solve over twenty load steps) are solve-dominated.}
\label{fig:timing}
\end{figure}

For each benchmark, we record time spent in each pipeline stage. Figure~\ref{fig:timing} shows the breakdown across the five physics classes. The orchestration cost is about 2.8 seconds and is nearly fixed across problems. It is dominated by starting a fresh Python subprocess for the FEniCS solver. Solve cost varies with physics. On small benchmark meshes with sub-second solves, orchestration is most of the total time. On practical problems, where solves take seconds to minutes, the fixed overhead becomes small. In the L-bracket demonstration, which takes 87.5 seconds end to end, the orchestration fraction is below 4\%.

\subsection{Scaling}\label{sec:scaling}

Table~\ref{tab:scaling} shows element counts and solve times for representative cases. Element counts range from a few hundred for small two-dimensional cases to 37,000 for the finest three-dimensional rubber block. Solve times range from below 0.1 seconds for small linear cases to about 40 seconds for the L-bracket elastoplastic solve. The system handles this range without changing the pipeline. Some cases do not have logged solve times because they were run only as part of mesh-convergence sweeps.

\begin{table}[h!]
\centering
\caption{Element counts and solve times for a representative subset of the benchmark suite. Em-dashes indicate cases run as part of mesh-convergence sweeps where solve time was not separately logged.}
\label{tab:scaling}
\small
\renewcommand{\arraystretch}{1.15}
\begin{tabular}{lrr}
\toprule
Problem & Elements & Solve (s) \\
\midrule
Thermo-Mechanical (2D) & 208 & 0.06 \\
Plasticity (3D bar, J2) & 254 & 9.62 \\
Cook's membrane (CG2 medium) & 406 & --- \\
L-bracket (LLM geom.) & 790 & 39.44 \\
Plate w/ hole (linear) & 946 & 0.07 \\
Thermo-Mechanical (3D) & 1{,}024 & --- \\
Hyperelasticity (3D box) & 1{,}367 & 0.05 \\
Phase-Field SENT (2D) & 6{,}989 & 11.96 \\
Phase-Field SENT (3D slab) & 8{,}444 & --- \\
Cantilever (3D, finest) & 12{,}832 & --- \\
Rubber block (3D) & 37{,}148 & --- \\
\bottomrule
\end{tabular}
\end{table}

\section{Robustness and Limitations}\label{sec:robustness}

\subsection{Reproducibility}\label{sec:reproducibility}

We re-ran three benchmark cases five times each from the same natural-language prompt at temperature zero: the 3D linear elastic cantilever, the 3D bar elastoplasticity case, and the 2D SENT phase-field fracture case. Table~\ref{tab:reproducibility} reports the mean and standard deviation of scalar metrics.

For the true 3D cantilever validation case, reproducibility was evaluated on the finest saved validation configuration (L5: $h_{\max}=2.5$ mm, $h_{\min}=0.8$ mm, second-order tetrahedra). Over five repeated runs, the tip displacement and the saved bending-stress probe value $\sigma_{xx}(x=20~\text{mm}, y=20~\text{mm}, z=5~\text{mm})$ were bit-identical across runs. Solve time varies because of background system load, not because of changes in the generated specification or template. The 2D SENT phase-field case likewise produced bit-identical physical scalars across runs.

The elastoplastic case shows small physical variability. Maximum displacement has a standard deviation of about 0.33 mm on a mean of 12.85 mm, or about 2.6\%. Maximum stress has a standard deviation of about 10 MPa on a mean of 390 MPa, also about 2.6\%. The parsed specification and generated FEniCS script are identical across runs. The observed spread appears in the nonlinear elastoplastic solve and may reflect numerical sensitivity in the path-dependent return-mapping/update sequence; its exact source was not isolated here. This spread is small relative to the scale of the response, but it is not zero.

\begin{table}[h!]
\centering
\caption{Reproducibility across five repeated runs from the same natural-language prompt. The 3D cantilever row corresponds to the true cantilever validation case at the finest saved validation configuration (L5). The cantilever and phase-field cases produce bit-identical physical scalars, while the elastoplastic case shows about 2.6\% relative spread on displacement and stress.}
\label{tab:reproducibility}
\small
\renewcommand{\arraystretch}{1.15}
\begin{tabular}{llrr}
\toprule
Benchmark & Metric & Mean & Std. dev. \\
\midrule
\multirow{3}{*}{3D Cantilever} & disp. (mm) & 0.38317 & 0.000 \\
                                & $\sigma_{xx}$ probe (MPa) & 53.994 & 0.000 \\
                                & solve time (s) & 4.783 & 1.810 \\
\midrule
\multirow{2}{*}{3D Plasticity} & disp. (mm) & 12.849 & 0.328 \\
                                & stress (MPa) & 390.34 & 10.20 \\
\midrule
2D SENT (AT2) & disp. (mm) & 0.006149 & 0.000 \\
\bottomrule
\end{tabular}
\end{table}

\subsection{Prompt Sensitivity}\label{sec:prompt_sensitivity}

For the L-bracket case, we wrote three prompt variants and ran each through the full pipeline. Table~\ref{tab:prompt_sensitivity} summarizes the results.

The minimal variant omits material properties, the fillet, and most boundary-condition details. The parser cannot extract enough information to produce a valid specification, and the validator rejects it. This is intended. The system should not silently invent a material model and constraints for a sparse prompt. The moderate variant used in Section~\ref{sec:lbracket} and the verbose variant both parse successfully and produce bit-identical results: the same final maximum displacement magnitude and maximum stress. The verbose version includes redundant engineering context that the parser ignores once the required physics is specified. This table reports the top-level maximum displacement magnitude; the force-displacement plot in Figure~\ref{fig:lbracket} uses the componentwise displacement history from the load steps.

\begin{table}[h!]
\centering
\caption{Prompt sensitivity for the 3D L-bracket case across three prompt variants. The minimal variant fails at validation because too much information is missing. The moderate and verbose variants produce bit-identical results.}
\label{tab:prompt_sensitivity}
\small
\renewcommand{\arraystretch}{1.15}
\begin{tabular}{lcrr}
\toprule
Variant & Parse OK & Final max disp. mag. (mm) & Stress (MPa) \\
\midrule
Minimal & No & --- & --- \\
Moderate & Yes & 50.782 & 493.00 \\
Verbose & Yes & 50.782 & 493.00 \\
\bottomrule
\end{tabular}
\end{table}

\subsection{Failure Modes}\label{sec:failure_modes}

The main parser-side failure mode is under-specification. The parser can infer common material or problem-class defaults, but it cannot rescue prompts that omit essential boundary conditions, loading, or material behavior. The parser benchmark shows that validation and retry repair many first-pass failures, but the system does not guarantee correctness for arbitrary natural language.

The main remaining geometry failure mode in the current benchmark is invalid geometry. Boundary-tagging and mesh-failure errors did not occur in this 10-case run, but one harder obround-slot case still failed after three attempts. Incorrect construction of composite features after boolean operations remains the most important geometry risk. Worked examples, stricter boundary-tagging instructions, targeted pre-execution guards, and validation checks help, but custom geometry generation is still not a substitute for a general CAD system.

Stress extraction near sharp stress concentrations is sensitive to projection. A global $L^2$ projection of von Mises stress smears the peak by 10 to 15\% even on adequate meshes. Direct evaluation of stress tensor components followed by analytical von Mises computation matches the Kirsch reference. This is a standard finite element issue, but it is easy to miss in automated pipelines.

Convergence failures in elastoplasticity at high load are recovered by reducing the load step size on retry. The original Picard scheme diverged after significant yielding because the frozen elastic stiffness was too far from the algorithmic tangent. The current initial-stiffness Newton scheme with residual-driven correction is more stable, but it is still less efficient than a full consistent-tangent implementation.

Volumetric locking in nearly incompressible materials with displacement-only elements remains a limitation. The Cook's membrane benchmark uses compressible parameters to avoid this issue. A user who specifies a Poisson ratio above 0.49 triggers a validator warning, but the current solve still proceeds with displacement-only elements. This can produce large stiffness errors in bending-dominated problems.

\subsection{Limitations}\label{sec:limitations}

The geometry stage handles standard shapes through deterministic generators and some novel shapes through LLM-generated Gmsh code, but it cannot match dedicated CAD tools for complex industrial parts. For such parts, the practical route is to import a STEP file from external CAD software. The pipeline can mesh and analyze imported geometry, but that is outside a pure natural-language-only workflow.

Full topology optimization is not supported. The current optimization mode is gradient-free design optimization over low-dimensional parameter spaces. Integration with dolfin-adjoint would enable gradient-based optimization for differentiable subproblems, but would require special handling for path-dependent and staggered physics.

Contact mechanics is not implemented. This is a major limitation for many engineering applications.

Mesh adaptivity is heuristic rather than error-driven. The mesher refines based on geometric features, boundary conditions, and known crack-path regions. It does not use a posteriori error estimates from the solution. This is adequate for the benchmarks here, but it may miss interior features that are not known before the solve.

The quality of parsing and custom geometry generation depends on the underlying LLM. The architecture is model-swappable through one wrapper, but smaller models may fail on complex multi-physics prompts or non-catalog geometry. The deterministic solver layer is independent of the model, but the front-end success rate is not.

\section{Discussion}\label{sec:discussion}

The main result is not that an LLM can write mechanics solvers. It is that a natural-language interface can be placed above deterministic FEniCS templates, with the LLM kept away from numerically sensitive code. This boundary changes what must be validated. The deterministic template layer must be checked against mechanics references. The LLM-facing layer must be checked for parsing, repair, geometry generation, and failure modes.

The mechanics validation shows that the deterministic templates solve the intended benchmark problems across five physics classes. Linear elasticity, smooth hyperelasticity, and thermo-mechanical coupling reach sub-percent or near-machine-precision agreement where analytical references are available. Elastoplasticity and phase-field fracture have larger but explainable errors in the 2--5\% range. These results support the breadth claim for the deterministic backend.

The parser benchmark supports a different claim. It shows that the structured-specification front end works well under validation and repair. First-pass valid parsing is not perfect: 40.0\% of the benchmark cases require repair. After retry, all 15 cases produce valid specifications, with perfect problem-class accuracy and one output-field omission. The correct interpretation is not that the parser is a perfect clean rejector. It is that the parser, validator, and retry loop form a useful repair-oriented specification interface.

The geometry benchmark gives 90\% success on this 10-case set. Retry did not recover the remaining case. The only unrecovered failure is an invalid-geometry failure in a harder obround-slot case. This is not evidence of broad reliability for arbitrary CAD-like parts. The L-bracket example shows that the path can support a non-catalog geometry when it succeeds; the benchmark gives the aggregate reliability result.

This framing addresses the main mismatch that can occur in papers on LLM-driven FEA. If the paper claims to be about an LLM interface but only reports mechanics validation, the evidence does not match the claim. If the paper reports only parser metrics without validating the mechanics layer, the simulation results are ungrounded. The present architecture needs both. The backend validation shows that the deterministic templates are correct for the supported physics. The front-end benchmarks show how well the LLM-facing layer produces the specifications and geometries that feed those templates.

Compared with ALL-FEM and MCP-SIM, the difference is not that this system is more general. It is more constrained. ALL-FEM and MCP-SIM allow the LLM to generate FEniCS code and then manage the resulting reliability problem through fine-tuning, multi-agent correction, or execution feedback. Our system removes that failure mode by construction. The cost is manual template development for each new physics class. The benefit is that the solver path is deterministic, inspectable, and independent of LLM code-generation quality.

The system is therefore best described as a constrained natural-language interface to variational simulation. It is not an autonomous mechanics agent. It does not discover weak forms. It does not guarantee correctness for arbitrary prompts. It provides a controlled route from natural language to a validated specification, then to deterministic FEniCS templates, with measured front-end behavior and measured backend accuracy.

\section{Conclusion}\label{sec:conclusion}

We presented a constrained natural-language interface for multi-physics simulations in FEniCS. The LLM parses prompts into structured specifications, generates Gmsh code only for custom non-catalog geometries, and uses error feedback for those front-end tasks. It does not write FEniCS templates, derive weak forms, or generate solver code. Validated specifications are routed to five human-written UFL templates covering linear elasticity, hyperelasticity, elastoplasticity, thermo-mechanical coupling, and phase-field fracture.

The deterministic mechanics layer was validated against analytical and published references in two and three dimensions. Smooth linear and nonlinear cases reached sub-percent agreement, while elastoplasticity and phase-field fracture reached the 2--5\% range. The same architecture also supports parametric sweeps and low-dimensional bounded gradient-free design optimization.

The LLM-facing front end was evaluated separately. In a 15-prompt parser benchmark, first-pass valid parsing was 60.0\%, retry repaired the remaining 40.0\%, and the final valid parse rate was 100.0\%. Problem-class accuracy was 100.0\%, and field-extraction accuracy was 97.1\%. In a 10-case custom geometry benchmark, first-pass success was 90.0\%, retry recovery was 0.0\%, and final success was 90.0\%. The one unrecovered case was an invalid-geometry failure in a harder obround-slot geometry. These results show that the parser is effective on this benchmark under repair and that the current constrained prompting, boundary-tagging instructions, and validation guards are effective on this benchmark.

The L-bracket case demonstrates the full pipeline on one successful non-catalog geometry. The system generates a filleted, holed 3D bracket through the LLM-to-Gmsh path and runs a forty-step elastoplastic analysis from one prompt in 87.5 seconds. This demonstration shows what the interface can do on a successful case, while the geometry benchmark gives the reliability picture.

The main architectural claim is that removing the LLM from the solver path avoids a major failure mode of direct FEniCS code generation. The cost is that each new physics class requires a human-written template. The benefit is that the numerical method remains deterministic and inspectable. Future work includes contact mechanics, mixed formulations for nearly incompressible materials, error-driven adaptive mesh refinement, gradient-based optimization for differentiable subproblems, and broader parser and geometry benchmarks across multiple LLM backends.

\bibliographystyle{unsrtnat}
\bibliography{bibliography.bib}

\onecolumn
\setcounter{figure}{0}
\setcounter{table}{0}
\renewcommand{\thefigure}{S\arabic{figure}}
\renewcommand{\thetable}{S\arabic{table}}

\phantomsection
\section*{Supplementary Information}
\label{sec:si}

\subsection*{Parser benchmark prompt list}

The supplementary parser prompt list for the paper-facing set contains the 15 parser benchmark prompts only: 10 accept cases spanning the five physics classes used in the paper, together with 2 ambiguous prompts and 3 invalid prompts.

\begingroup
\small
\setlength{\tabcolsep}{5pt}
\renewcommand{\arraystretch}{1.20}
\begin{longtable}{@{}>{\raggedright\arraybackslash}p{0.28\textwidth}
                  >{\raggedright\arraybackslash}p{0.11\textwidth}
                  >{\raggedright\arraybackslash}p{0.19\textwidth}
                  >{\raggedright\arraybackslash}p{0.34\textwidth}@{}}
\caption{Parser benchmark prompts used in the paper-facing benchmark set.}
\label{tab:si_parser_prompts}\\
\toprule
\textbf{Case ID} & \textbf{Category} & \textbf{Target class} & \textbf{Prompt summary} \\
\midrule
\endfirsthead

\toprule
\textbf{Case ID} & \textbf{Category} & \textbf{Target class} & \textbf{Prompt summary} \\
\midrule
\endhead

\bottomrule
\endfoot

\nolinkurl{parser_01_linear_cantilever_informal} & \texttt{accept} & \nolinkurl{linear_elasticity} & Steel cantilever, 250 mm long and 25 mm wide, left end clamped, 750 N downward tip load. \\
\nolinkurl{parser_02_linear_mixed_units_plate_hole} & \texttt{accept} & \nolinkurl{linear_elasticity} & Aluminum plate with mixed units, centered 6 mm hole, left edge clamped, 50 MPa tension on the right. \\
\nolinkurl{parser_03_hyper_block_shear} & \texttt{accept} & \nolinkurl{hyperelasticity} & 3D rubber block, bottom fixed, top pushed 5 mm in shear with large deformation expected. \\
\nolinkurl{parser_04_hyper_silicone_gasket} & \texttt{accept} & \nolinkurl{hyperelasticity} & Silicone gasket compressed 25\% between rigid plates. \\
\nolinkurl{parser_05_plastic_coupon_displacement} & \texttt{accept} & \nolinkurl{elastoplasticity} & Steel coupon pulled by prescribed 0.6 mm displacement past yield, with hardening. \\
\nolinkurl{parser_06_plastic_notched_strip} & \texttt{accept} & \nolinkurl{elastoplasticity} & Aluminum notched strip under 12 kN tensile loading to check permanent set. \\
\nolinkurl{parser_07_thermo_copper_bar} & \texttt{accept} & \nolinkurl{thermo_mechanical} & Copper bar with left edge pinned and a 180\,$^\circ$C to 40\,$^\circ$C temperature gradient. \\
\nolinkurl{parser_08_thermo_block_mixed_units} & \texttt{accept} & \nolinkurl{thermo_mechanical} & Steel block in mixed units, bottom fixed, one end at 200\,$^\circ$C and the other at 50\,$^\circ$C. \\
\nolinkurl{parser_09_phase_field_sent} & \texttt{accept} & \nolinkurl{phase_field_fracture} & 2D single-edge-notch tension plate pulled upward until fracture. \\
\nolinkurl{parser_10_phase_field_bending} & \texttt{accept} & \nolinkurl{phase_field_fracture} & Notched concrete beam in bending with midspan point load and crack propagation study. \\
\nolinkurl{parser_13_ambiguous_too_short} & \texttt{ambiguous} & --- & One-word prompt: ``Beam.'' \\
\nolinkurl{parser_14_ambiguous_vague} & \texttt{ambiguous} & --- & Vague metal-part prompt with approximate length and unspecified force location. \\
\nolinkurl{parser_15_invalid_oversized_hole} & \texttt{invalid} & --- & Steel plate with an oversized 60 mm radius hole in a 100 mm by 100 mm plate. \\
\nolinkurl{parser_16_invalid_bad_material} & \texttt{invalid} & --- & Rubber block with unphysical material parameters: Poisson ratio 0.8 and negative modulus. \\
\nolinkurl{parser_17_invalid_free_body} & \texttt{invalid} & --- & Steel cube with no supports, no clamps, and no loads. \\
\end{longtable}
\endgroup

\subsection*{Custom-geometry benchmark prompt list}

\noindent 
Table~\ref{tab:si_geometry_prompts} lists the custom-geometry benchmark prompts used for the real LLM-to-Gmsh path. It identifies the target geometry, the required named boundaries, and the one remaining hard failure case, \nolinkurl{geom_10_obround_slot_plate}, from the final benchmark.

\begingroup
\small
\setlength{\tabcolsep}{5pt}
\renewcommand{\arraystretch}{1.20}
\begin{longtable}{@{}>{\raggedright\arraybackslash}p{0.26\textwidth}
                  >{\raggedright\arraybackslash}p{0.18\textwidth}
                  >{\raggedright\arraybackslash}p{0.35\textwidth}
                  >{\raggedright\arraybackslash}p{0.15\textwidth}@{}}
\caption{Custom-geometry benchmark prompts used for the real LLM-to-Gmsh path.}
\label{tab:si_geometry_prompts}\\
\toprule
\textbf{Case ID} & \textbf{Geometry type} & \textbf{Geometry summary} & \textbf{Required boundary names} \\
\midrule
\endfirsthead

\toprule
\textbf{Case ID} & \textbf{Geometry type} & \textbf{Geometry summary} & \textbf{Required boundary names} \\
\midrule
\endhead

\bottomrule
\endfoot

\nolinkurl{geom_01_box_through_hole} & \nolinkurl{simple_cut} & Rectangular block with a centered through-hole along the $z$-axis. & \nolinkurl{fixed_face}; \nolinkurl{loaded_face} \\
\nolinkurl{geom_02_stepped_block_notch} & \nolinkurl{boolean_fuse_cut} & Stepped solid formed by fusing two boxes, then cutting a rectangular side notch. & \nolinkurl{support_face}; \nolinkurl{traction_face} \\
\nolinkurl{geom_03_l_bracket_fillet_hole} & \nolinkurl{fillet_hole} & L-bracket formed from two fused arms, with an inner fillet and a through-hole. & \nolinkurl{fixed_face}; \nolinkurl{top_face} \\
\nolinkurl{geom_04_t_joint_two_holes} & \nolinkurl{multi_boolean} & T-shaped bracket with two through-holes in the flange. & \nolinkurl{stem_base_face}; \nolinkurl{flange_top_face} \\
\nolinkurl{geom_05_mounting_tab_awkward} & \nolinkurl{awkward_prompt} & Mounting-tab style block with a corner cut and one through-hole. & \nolinkurl{left_face}; \nolinkurl{right_face} \\
\nolinkurl{geom_06_offset_boss_plate} & \nolinkurl{awkward_fuse_cut} & Base plate with an offset cylindrical boss and a through-hole through the boss. & \nolinkurl{plate_bottom_face}; \nolinkurl{boss_top_face} \\
\nolinkurl{geom_07_slotted_plate_two_bolts} & \nolinkurl{slot_hole_combo} & Flat bracket with a centered through-slot and two bolt holes. & \nolinkurl{left_face}; \nolinkurl{right_face} \\
\nolinkurl{geom_08_channel_cut_block} & \nolinkurl{channel_cut} & Solid block with a centered U-shaped top channel, leaving a bottom flange and side walls. & \nolinkurl{left_end_face}; \nolinkurl{right_end_face} \\
\nolinkurl{geom_09_clevis_cross_hole_block} & \nolinkurl{cross_hole_clevis} & Clevis-like block with a front slot and a cross-hole through the two ears. & \nolinkurl{back_face}; \nolinkurl{fork_tip_faces} \\
\nolinkurl{geom_10_obround_slot_plate} & \nolinkurl{difficult_obround_slot} & Plate with an obround through-slot and two through-holes. This remained the unrecovered benchmark failure. & \nolinkurl{left_edge_face}; \nolinkurl{right_edge_face} \\
\end{longtable}
\endgroup

\subsection*{Parser problem-class confusion matrix}

\noindent 
The confusion matrix reported here is restricted to the 10 labeled accept cases in the five physics classes used in the paper.

\begin{table}[h!]
\centering
\caption{Full parser confusion matrix for the five physics classes used in the paper. Abbreviations: LE = linear elasticity, HE = hyperelasticity, EP = elastoplasticity, TM = thermo-mechanical, PF = phase-field fracture, NP = no prediction.}
\label{tab:si_confusion}
\small
\renewcommand{\arraystretch}{1.15}
\begin{tabular}{lrrrrrr}
\toprule
\textbf{True class} & \textbf{LE} & \textbf{HE} & \textbf{EP} & \textbf{TM} & \textbf{PF} & \textbf{NP} \\
\midrule
\nolinkurl{linear_elasticity}     & 2 & 0 & 0 & 0 & 0 & 0 \\
\nolinkurl{hyperelasticity}        & 0 & 2 & 0 & 0 & 0 & 0 \\
\nolinkurl{elastoplasticity}       & 0 & 0 & 2 & 0 & 0 & 0 \\
\nolinkurl{thermo_mechanical}     & 0 & 0 & 0 & 2 & 0 & 0 \\
\nolinkurl{phase_field_fracture} & 0 & 0 & 0 & 0 & 2 & 0 \\
\bottomrule
\end{tabular}
\end{table}

\subsection*{Example Parsed JSON Specification}
\noindent 
To illustrate the abstraction boundary between the LLM and the deterministic solver, the listing below shows a representative parsed JSON specification. This example corresponds to the three-dimensional linear elastic cantilever benchmark. The LLM populates these specific, constrained fields based on the natural-language prompt, and the deterministic template dispatcher uses this structured dictionary to assemble the FEniCS UFL script. 
\newpage
\begin{verbatim}
{
  "problem_metadata": {
    "problem_class": "linear_elasticity",
    "dimension": 3,
    "analysis_type": "static"
  },
  "geometry": {
    "shape_type": "box",
    "parameters": {
      "length": 200.0,
      "width": 20.0,
      "height": 10.0
    },
    "units": "mm-N-MPa"
  },
  "material": {
    "name": "steel",
    "constitutive_model": "linear_elastic",
    "properties": {
      "E": 210000.0,
      "nu": 0.3
    }
  },
  "boundary_conditions": [
    {
      "type": "dirichlet",
      "location": "left_face",
      "constrained_dofs": ["all"],
      "value": 0.0
    }
  ],
  "loads": [
    {
      "type": "distributed",
      "location": "right_face",
      "magnitude": 1.0,
      "direction": "-y"
    }
  ],
  "mesh": {
    "density": "coarse",
    "element_order": 2,
    "h_max": 2.5,
    "h_min": 0.8
  },
  "solver": {},
  "output": {
    "requested_fields": [
      "displacement",
      "von_mises_stress"
    ]}}
\end{verbatim}

\end{document}